\documentclass[pdflatex,sn-mathphys-num]{sn-jnl}

\usepackage{graphicx}%
\usepackage{multirow}%
\usepackage{amsmath,amssymb,amsfonts}%
\usepackage{amsthm}%
\usepackage{mathrsfs}%
\usepackage[title]{appendix}%
\usepackage{xcolor}%
\usepackage{textcomp}%
\usepackage{tabularx}
\usepackage{manyfoot}%
\usepackage{booktabs}%
\usepackage{algorithm}%
\usepackage{algorithmicx}%
\usepackage{algpseudocode}%
\usepackage{listings}%
\usepackage{lineno}
\usepackage{siunitx}
\usepackage{amsmath}
\usepackage{booktabs} 
\usepackage{booktabs}   
\usepackage{rotating}   
\usepackage{caption}    
\usepackage{caption}
\usepackage{threeparttable}
\usepackage[T1]{fontenc}   
\usepackage{lmodern}       

\usepackage{mathrsfs}
\DeclareFontFamily{U}{rsfs}{}
\DeclareFontShape{U}{rsfs}{m}{n}{<->rsfs10}{}
\DeclareFontShape{U}{rsfs}{m}{sl}{<->rsfsl10}{}
\DeclareFontShape{U}{rsfs}{bx}{n}{<->rsfs10}{}
\DeclareFontShape{OMS}{cmsy}{m}{n}{<-> s*[1.0] cmsy10}{}

\usepackage{array}      
\usepackage{booktabs}   
\usepackage{rotating}   
\newcolumntype{C}[1]{>{\centering\arraybackslash}p{#1}}

\theoremstyle{plain}
\theoremstyle{plain}%

\theoremstyle{plain}%

\begin{document}


\title[Article Title]{PdNeuRAM: Forming-Free, Multi-Bit Pd/HfO\textsubscript{2} ReRAM for Energy-Efficient Computing}


\author[1]{\fnm{Erbing} \sur{Hua}}\email{e.hua@tudelft.nl}
\author[1]{\fnm{Theofilos} \sur{Spyrou}}\email{t.spyrou@tudelft.nl}
\author[2,3]{\fnm{Majid} \sur{Ahmadi}}\email{majid.ahmadi@rug.nl}
\author[4]{\fnm{Abdul Momin} \sur{Syed}}\email{abdulmomin.syed@kaust.edu.sa}
\author[1]{\fnm{Hanzhi} \sur{Xun}}\email{h.xun@tudelft.nl}
\author[5]{\fnm{Laurentiu Braic} }\email{laurentiu.braic@kaust.edu.sa}
\author[2]{\fnm{Ewout} \sur{van der Veer}}\email{ewout.van.der.veer@rug.nl}
\author[4]{\fnm{Nazek} \sur{Elatab}}\email{nazek.elatab@kaust.edu.sa}
\author[1]{\fnm{Anteneh} \sur{Gebregiorgis}}\email{a.b.gebregiorgis@tudelft.nl}
\author[1]{\fnm{Georgi} \sur{Gaydadjiev}}\email{g.n.gaydadjiev@tudelft.nl} 
\author[2,3]{\fnm{Beatriz} \sur{Noheda}}\email{b.noheda@rug.nl}
\author[1]{\fnm{Said} \sur{Hamdioui}}\email{s.hamdioui@tudelft.nl}
\author[1]{\fnm{Ryoichi} \sur{Ishihara}}\email{r.ishihara@tudelft.nl}
\author*[1]{\fnm{Heba} \sur{Abunahla}}\email{h.n.abunahla@tudelft.nl}

\affil[1]{\orgdiv{Department of Quantum and Computer Engineering}, \orgname{Delft University of Technology}, \orgaddress{\city{Delft}, \country{The Netherlands}}}

\affil[2]{\orgdiv{Zernike Institute for Advanced Materials}, \orgname{University of Groningen}, \orgaddress{\city{Groningen}, \country{The Netherlands}}}

\affil[3]{\orgdiv{CogniGron center}, \orgname{University of Groningen}, \orgaddress{\city{Groningen}, \country{The Netherlands}}}

\affil[4]{\orgdiv{Computer Electrical Mathematical Science and Engineering Division, Electrical and Computer Engineering}, \orgname{King Abdullah University of Science and Technology}, \orgaddress{\city{Thuwal}, \country{Saudi Arabia}}}
\affil[5]{\orgdiv{Core Labs}, \orgname{King Abdullah University of Science and Technology}, \orgaddress{\city{Thuwal}, \country{Saudi Arabia}}}


\abstract{
Memristor technology shows great promise for energy-efficient computing; yet it grapples with challenges like resistance drift, and inherent variability. For filamentary Resistive RAM (\textit{ReRAM}), one of the most investigated types of memristive devices, the expensive electroforming step required to create conductive pathways, results in increased power/area overheads and reduced endurance. In this study, we present novel HfO\textsubscript{2}-based forming-free \textit{ReRAM} devices, PdNeuRAM, that operate at low voltages, support multi-bit functionality, and display reduced variability. Through a deep understanding and comprehensive material characterization, we discover the key process that allows this unique behavior: a Pd–O–Hf configuration that capitalizes on Pd innate affinity for integratinging into HfO\textsubscript{2-x}. This structure actively facilitates charge redistribution at room temperature, effectively eliminating the need for electroforming. Moreover, the fabricated \textit{ReRAM} device provides tunable resistance states for dense memory and reduces programming and reading energy by $43\%$ and $73\%$, respectively using  spiking neural networks (\textit{SNNs}). This study reveals novel mechanistic insights and delineates a strategic roadmap for the realization of power-efficient and cost-effective \textit{ReRAM} devices.
}




\maketitle



In this emerging era defined by Artificial General Intelligence (\textit{AGI}) and Internet of Things (\textit{IoT}) technologies~\cite{kudithipudi2025neuromorphic}, Computing-in-Memory (\textit{CIM}), where computation and storage seamlessly converge in a single physical locale, has ascended as a visionary next-generation computing paradigm~\cite{hamdioui2015memristor}. Among the many candidate technologies for computing cells, \textit{ReRAM} has captivated both academia and industry with its compelling attributes, including non-volatility, zero leakage power, a simplified structure, high integration density, rapid switching speeds, and inherent \textit{CMOS} compatibility~\cite{kudithipudi2025neuromorphic,chen2016ultrathin,wang2022forming}. However, beneath these luminous advantages lie formidable challenges, most notably, the prerequisite for electroforming pre-processing, where the electroforming phase necessitates the application of substantially elevated voltages to establish Conductive Filaments (\textit{CF}) within the resistive medium~\cite{wang2022forming}. It not only intensifies energy consumption and complicates fabrication but also compromises device reliability and endurance~\cite{tsai2015structure, park2024multi, chen2010ultrathin}. Moreover, the pronounced current density during \textit{ReRAM} write operations further exacerbates concerns about power consumption and long-term durability, often compelling designers to enlarge \textit{ReRAM} cell dimensions, a trade-off that diminishes circuit packing density and inflates manufacturing costs~\cite{tsai2015structure, huang2016rram}.

To address these challenges, HfO\textsubscript{2} has been widely explored as a premier dielectric candidate, celebrated for its superior breakdown voltage, robust thermal stability, and exceptional \textit{CMOS} compatibility. These remarkable attributes promise not only a reduction in operational power consumption but also enhanced thermal performance over alternative dielectrics~\cite{stecconi2022filamentary,wedig2016nanoscale,nukala2021reversible,falcone2025all}. Furthermore, the incorporation of Multi-Level Resistance (\textit{MLR}) technology in HfO\textsubscript{2} boosts memory density and slashes manufacturing costs~\cite{beckmann2016nanoscale, nakayama2007pulse, xu2013understanding}, offering significant advantages for neuromorphic computing applications. Yet, current specialized fabrication methods, such as thermal annealing~\cite{ding2023forming}, X-ray irradiation~\cite{wang2022forming}, exotic element doping~\cite{ding2023forming,kim2016forming}, and plasma treatments~\cite{wu2023reversing,kim2016forming}, are designed to introduce defect states and reduce forming voltages. Aforementioned techniques will increase manufacturing costs and complexity, typically demand substantial energy, and may inadvertently impair transistor performance~\cite{ikraiam2025investigating}. Therefore, there is a compelling need for simpler, energy-efficient fabrication processes for forming-free \textit{ReRAM} devices.

\begin{figure*}[!t]
\centerline{\includegraphics[width=0.8\linewidth]{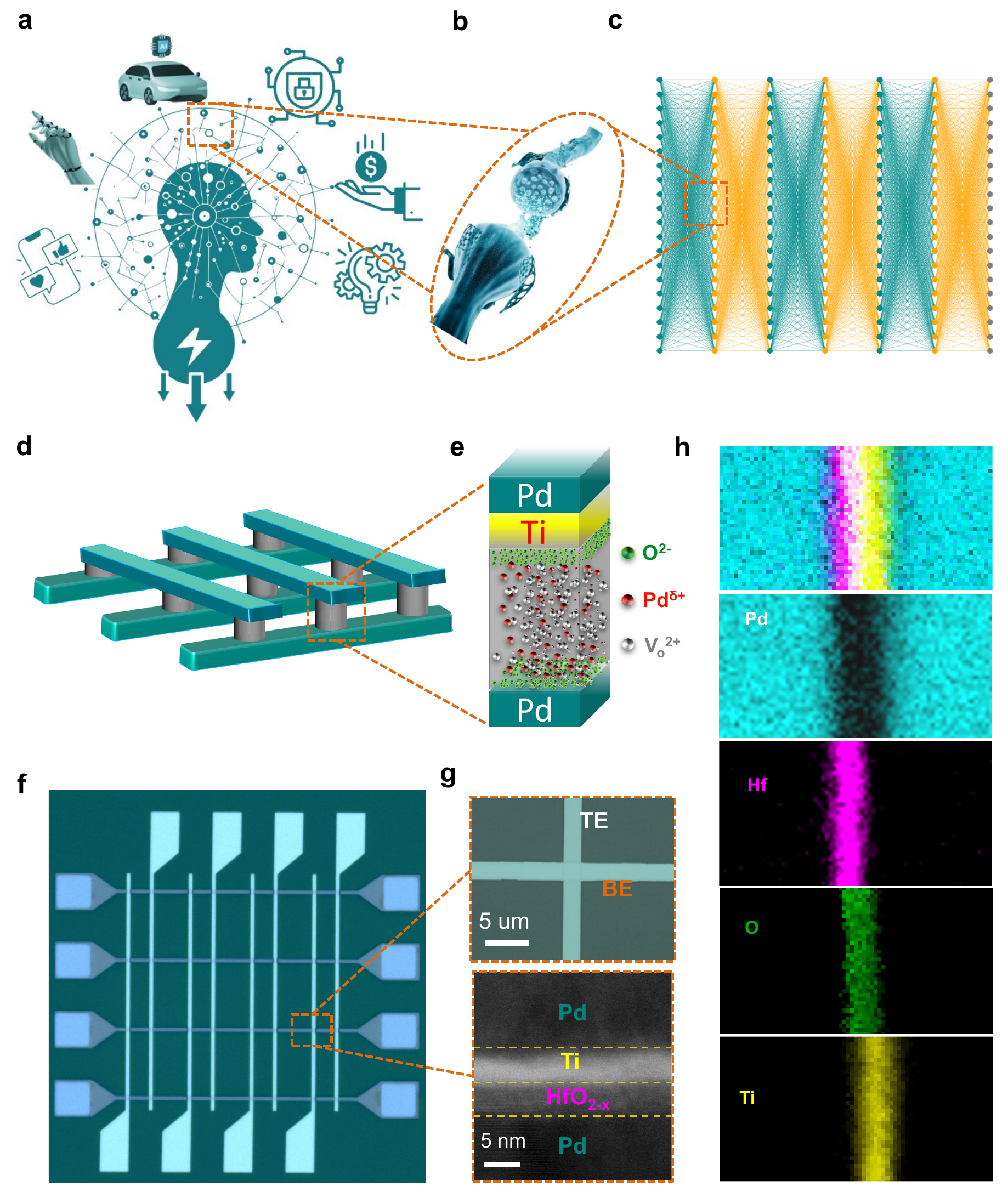}}
\caption{\textbar\quad \textbf{Overview of neuromorphic systems and \textit{ReRAM} device stacks.} \textbf{a} An overview of potential applications for energy-efficient neuromorphic systems. \textbf{b} Schematics of a biological synapse. \textbf{c} Schematics of fully connected artificial neural networks. \textbf{d} Schematic illustration of crossbars. \textbf{e} Schematics of the designed \textit{ReRAM} cell. \textbf{f} \textit{SEM} image of fabricated 4 $\times$ 8 crossbars. \textbf{g} The optical image of one cell (top) and \textit{STEM} image of the cross section of one Ox\textit{ReRAM} cell (bottom). \textbf{h} Cross-sectional Elemental Energy Loss Spectroscopy (\textit{EELS}) mapping of the PdNeuRAM cell in Low Resistance States (\textit{LRS}) with the structure of  Pd/HfO\textsubscript{2-x}/Ti/Pd (denoted as PdHT, also PdNeuRAM).}
\label{RRAM_fab}
\vspace{-1em}%
\end{figure*}

In this study,  inspired by the remarkable efficiency of biological neural cells, as illustrated in Fig.~\ref{RRAM_fab}a, b, which underpin the acceleration of neuromorphic devices for fully connected artificial neural networks (Fig.~\ref{RRAM_fab}c)~\cite{kudithipudi2025neuromorphic}, 
we proposed a streamlined, forming-free \textit{ReRAM} fabrication approach that leverages conventional Integrated Circuit (\textit{IC}) manufacturing processes. Notably, the presence of Pd, serving as both an electron injection source and catalytic center in HfO\textsubscript{2-x}, fosters a particular atomic configuration that diminishes charge diffusion barriers~\cite{traore2015hfo}. Synergistically, the cooperative action of these integrated stacks culminates in forming-free behavior. Electrical characterization reveals significantly lower initial resistivity in pristine PdHT devices compared to conventional Pt/HfO\textsubscript{2-x}/Ti/Pt (PtHT) devices~\cite{formingvoltage_ptht2016}. Moreover, the PdHT devices exhibit stable multilevel resistance states and low-power switching capabilities, achieved through modulation of \textit{RESET} stopping voltages (V\textsubscript{\textit{RESET, stop}}), where the \textit{RESET}  refers to switching the memory device to its high-resistance state; the bias at the switching point is defined as V\textsubscript{\textit{RESET}}; and the sweeping stop voltage is defined as V\textsubscript{\textit{RESET, stop}}. These attributes render the proposed PdHT device a compelling candidate for energy-efficient neuromorphic computing applications. Experimental validations further confirm substantial energy savings during both training and inference phases of \textit{SNN}, demonstrating effective performance in tasks such as image classification and gesture recognition.


\section*{Results}\label{sec2}
\addcontentsline{toc}{section}{Results}

\subsubsection*{Device characterization}

We designed crossbar arrays of devices with  2~$\mu$m $\times$ 2~$\mu$m node size (see Method), as displayed in Figs.~\ref{RRAM_fab}d, e.  The successful realization of the designed layout and stack is validated by optical microscopy and High-Resolution Scanning Transmission Electron Microscopy (\textit{HRSTEM}), as depicted in Figs.~\ref{RRAM_fab}f, g. The distinct layers of the device stacks are clearly visible in the cross-sectional \textit{HRSTEM} image, as shown in Fig.~\ref{RRAM_fab}g, where the HfO\textsubscript{2-x} layer exhibits a thickness of approximately 5 nm, sandwiched by 5 nm layers of Ti and Pd electrodes. Furthermore, \textit{HRSTEM-EELS} imaging of PdHT  (Fig.~\ref{RRAM_fab}h) confirms the precise elemental distribution.

\begin{figure*}[!t]
\centerline{\includegraphics[width=1.0\linewidth]{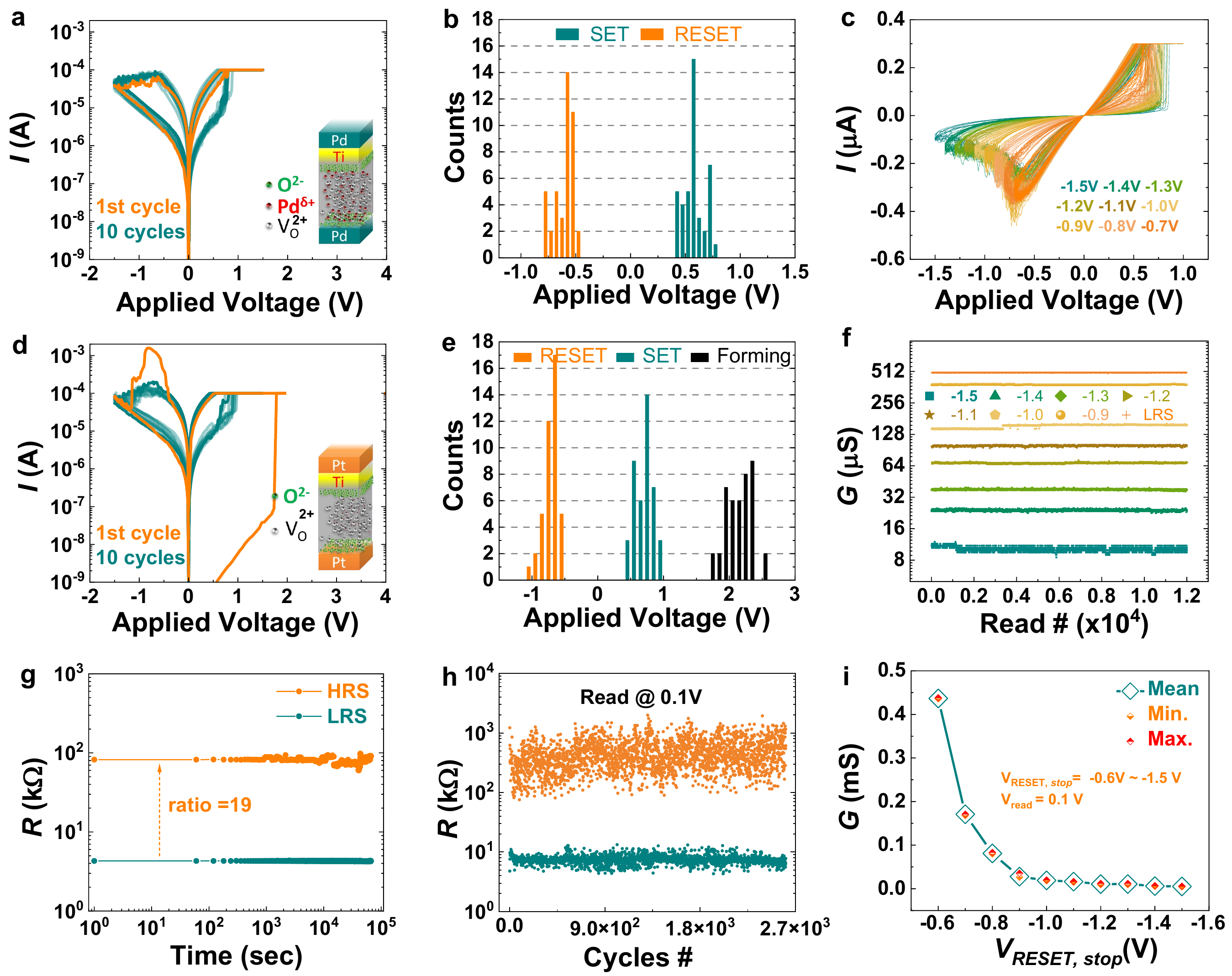}}
\vspace{0.0em}%
\caption{\textbar\quad\textbf{Electrical characterization of PdHT and PtHT memristors}. \textbf{a} I-V curves of forming-free PdHT memristors. The inset schematic diagrams show the PdHT structures. \textbf{b} Statistical analysis of V\textsubscript{\textit{SET}} and V\textsubscript{\textit{RESET}} for 42 PdHT devices. \textbf{c} I-V curves for tuning various V\textsubscript{RESET, stop}. \textbf{d}  I-V curves of conventional PtHT memristors. The inset schematic diagrams show the PtHT structures. \textbf{e} Statistical analysis of V\textsubscript{\textit{SET}}, V\textsubscript{\textit{RESET}} and V\textsubscript{\textit{Forming}} for PtHT memristors. \textbf{f} Reading stability test for 8 representative resistance states by 1.2 $\times$ 10\textsuperscript{4} readings under \SI{0.1}{V}. \textbf{g} Retention test of resistance states for 4.5 $\times$ 10\textsuperscript{4} seconds with reading intervals of 60 seconds. \textbf{h} Endurance test of PdHT by \textit{DC} sweeping over 2.5 $\times$ 10\textsuperscript{3} cycles and reading at \SI{0.1}{V}. \textbf{i} Conductance change over V\textsubscript{\textit{RESET, stop}} including its mean, maximum, and minimum conductance values as statistically analyzed in \textbf{f}.}
\label{IV}
\vspace{-1em}%
\end{figure*}

To examine the electrical properties of the fabricated devices, I–V characterizations were performed, as depicted in Fig.~\ref{IV}. Measurements on the electrical properties of the PdHT devices reveal three pivotal attributes for energy-efficient computing, including electroforming-free operation, low operating voltages, and tunable conductance. As illustrated in Fig.~\ref{IV}a, the PdHT devices manifest forming-free bipolar switching behavior. A statistical analysis of 42 randomly selected devices from the same die (Fig.~\ref{IV}b) reveals that the V\textsubscript{\textit{SET}} predominantly centers around \SI{0.56}{V}, while the V\textsubscript{\textit{RESET}} is near \SI{-0.58}{V}. For comparison, similar characterizations were conducted on PtHT devices fabricated using an identical process except for choosing Pt as the top and bottom electrodes (\textit{TE}/\textit{BE}). As shown in Fig.~\ref{IV}d, these PtHT devices show bipolar switching, which is similar to PdHT devices, but they require an additional electroforming step (as illustrated in Fig.~\ref{IV}e), a result consistent with previous reports~\cite{formingvoltage_ptht2016}. The Device-to-Device variability (\textit{D2D}) test (Fig.~\ref{IV}e) confirms that PtHT devices require an electroforming voltage of approximately \SI{2.3}{V}. Moreover, the statistical analysis presented in Table~\ref{tab:comparison} confirms a significant alleviation in variability and a reduction in both $V_{\textit{SET}}$ and $V_{\textit{RESET}}$ for the PdHT structure compared to both the PtHT devices and the state-of-the-art reports~\cite{ding2023forming,wang2022forming,kim2016forming}, as detailed in Table~\ref{tab:state-of-the-art}. Table~\ref{tab:state-of-the-art} presents a comprehensive comparison of state-of-the-art forming-free \textit{ReRAM} devices, encompassing both non-HfO\textsubscript{2} and HfO\textsubscript{2}-based systems. This analysis reveals that these devices typically exhibit one or more inherent limitations: they require elevated SET/RESET voltages, offer a constrained range of resistance states, or depend on specialized treatments to eliminate the conventional electroforming process. In contrast, our work successfully overcomes these issues, providing a more efficient and robust solution.

\begin{table}[ht]
    \centering
    \caption{\textbar\quad Comparison of operating voltages and their variability for PtHT and PdHT devices, including  cycle-to-cycle (\textit{C2C}) and \textit{D2D}.}
    \label{tab:comparison}
    \begin{tabular}{lcccc}
        \toprule
        \textbf{Structures} & \textbf{\textit{C2C (V\textsubscript{SET})}} & \textbf{\textit{D2D (V\textsubscript{SET})}} & \(\mathbf{V_{\textit{SET}}}\) & \(\mathbf{V_{\textit{RESET}}}\) \\
        \midrule
       PtHT & 11.0\% & 19.2\% & 0.68\,V & -0.72\,V \\
        PdHT & 7.7\% & 16.8\% & 0.56\,V & -0.58\,V \\
        \bottomrule
    \end{tabular}
\end{table}

Furthermore, the reliability of the PdHT devices was also studied and certified by further electrical measurements. The multibit capability of the PdHT devices was evaluated using a linear voltage sweeping mode. As shown in Fig.~\ref{IV}c, the resistance is continuously modulated by varying the V\textsubscript{\textit{RESET}}, thereby precisely controlling the rupture degree of the \textit{CF}. To ascertain the stability of the distinct resistance states, a \textit{READ}-number dependent stability test was conducted (Fig.~\ref{IV}f), which demonstrates analog state functionality that permits the storage of eight tunable multi-bit weights within a single memory cell. Furthermore, retention (Fig.~\ref{IV}g) and endurance (Fig.~\ref{IV}h) assessments reveal the  stable resistance states between \textit{LRS} and High Resistance States (\textit{\textit{HRS}}), with outstanding non-volatility, where the resistance states remain stable for over $4.5 \times 10\textsuperscript{4} $ seconds, and the device can endure 2.5 $\times$ 10\textsuperscript{3} cycles without noticeable performance degradation, though the performance is not presenting its optimum pattern (higher temperature of the retention test and higher number of cycles for the endurance test are needed for further investigation). Additionally, to gain deeper insight into the tunable resistance states, we conducted a comprehensive analysis of the conductance variation versus $V_{\textit{RESET, stop}}$, as depicted in Fig.~\ref{IV}i. The results reveal an exponential gradual decline in conductance with increasing $V_{\textit{RESET, stop}}$, indicating that the extent of \textit{CF} rupture can be modulated by adjusting the amplitude of $V_{\textit{RESET, stop}}$. This tunability may be further enhanced by controlling additional pulse parameters, such as pulse numbers, pulse amplitudes, and pulse width of $V_{\textit{RESET, stop}}$.


\begin{table*}[!ht]
\centering
\begin{threeparttable}
\centering
\fontsize{7.5pt}{11pt}\selectfont %
\setlength{\tabcolsep}{2pt} %
\renewcommand{\arraystretch}{1.3} %
\caption{\textbar\quad The-state-of-the-art forming-free devices}
\label{tab:state-of-the-art}
\begin{tabular}{%
  p{1.1cm}   
  p{1.4cm} 
  p{2.0cm} 
  p{1.0cm}   
  p{1.2cm} 
  p{1.5cm} 
  p{0.8cm} 
  p{1.0cm} 
  p{2.0cm} 
}
\toprule
\textbf{Types} & 
\textbf{TE} & 
\textbf{Th\textsubscript{\textit{OEL}}\footnotemark[1]} & 
\textbf{BE} & 
\textbf{On/Off} & 
\(\mathbf{V_{\textit{SET/RESET}}}\) & 
\textbf{MLR\footnotemark[2]} & 
\textbf{T\textsubscript{\textit{ann.}}\footnotemark[3]} & 
\textbf{Treatment} \\
\midrule
\multirow{5}{*}{\textbf{no HfO\(_2\)}} 
  & W~\cite{luo2018self}
  & 20\,nm WO\(_x\)
  & Pd
  & \(\sim20\)
  & 1.0,~-3.0
  & No
  & No
  & oxygen~plasma \\
& Ta~\cite{kim2016lowering}
  & 7\,nm Ta\(_2\)O\(_5\)
  & Pt
  & \(\ge10\)
  & 1.2,~-0.9
  & No
  & 600\(^\circ\)C
  & annealing \\
&SrTiO\(_3\)~\cite{wang2020electroforming}
  & 400\,nm MoO\(_3\)
  & Pt
  & \(\sim10\)
  & 3.5,~-3.8
  & No
  & 100\(^\circ\)C\
  & proton~injection \\
& Al~\cite{sun2019hybrid}
  & 150\,nm SiN\(_x\)
  & p\(^{++}\)\,Si
  & \(\sim10\)
  & 3.2,~-1.5
  & No
  & No
  & hydrogen~plasma \\
& TiN~\cite{chakrabarti2013multilevel}
  & 5\,nm HfTiO\(_x\)
  & TiN
  & \(\ge30\)
  & 1.0,~-1.0
  & 2
  & 400\(^\circ\)C\
  & annealing \\
\midrule

\multirow{3}{*}{\textbf{HfO\(_2\)}} 
  & Ge~\cite{ding2023forming}
  & 5\,nm HfO\(_2\)
  & Pd
  & \(\ge500\)
  & 3.2,~-0.8
  & No
  & 600\(^\circ\)C\
  & Ge~doping \\
& Al~\cite{kumar2020oxygen}
  & 60\,nm HfO\(_x\)
  & p\(^{++}\)\,Si
  & \(\sim3\)
  & 2.9,~-2.5
  & No
  & 550\(^\circ\)C\
  & elements~doping \\
& Pt~\cite{wang2022forming}
  & 20\,nm HfO\(_2\)
  & TiN
  & \(\ge10\)
  & 0.70,~-0.70
  & No
  & No
  & X-Ray\\
& TiN~\cite{chen2010ultrathin}
  & 3\,nm HfO\(_x\)
  & TiN
  & \(\sim10\)
  & 0.75,~-0.6
  & 2
  & 450\(^\circ\)C\
  & annealing \\
& Pd
  & 5\,nm HfO\textsubscript{2-x}
  & Pd
  & \(\sim19\)
  & 0.56,~-0.58
  & \(\ge3\)
  & No
  & No [This work] \\
\bottomrule
\end{tabular}
\begin{tablenotes}
  \item[1] Thickness of oxygen exchange layer.
  \item[2] Multi-level resistance and number of bits.
  \item[3] Temperature of thermal annealing
\end{tablenotes}
\end{threeparttable}
\end{table*}

\subsubsection*{Understanding the forming-free behavior}

\begin{figure*}[!t]
\centerline{\includegraphics[width=1.0\linewidth]{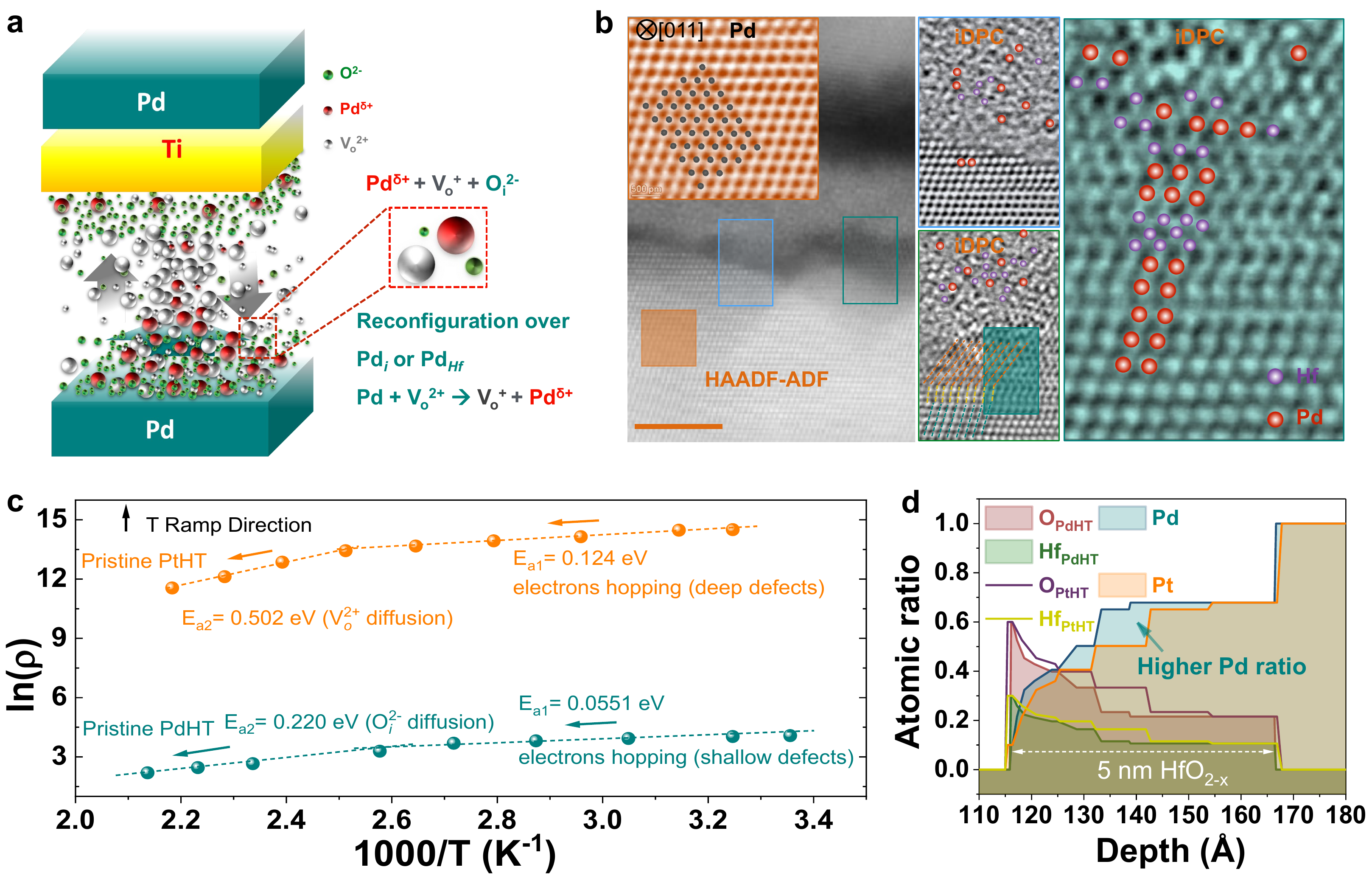}}
\vspace{-0.0em}%
\caption{\textbar\quad \textbf{Investigation of forming-free behavior} \textbf{a} schematic illustration of a forming-free mechanism. \textbf{b} \textit{STEM-HAADF} and\textit{ iDPC} images of the PdHT device. The scale bar is 5 nm. \textbf{c} Activation energy study for PtHT (orange) and PdHT (dark cyan). \textbf{d}  \textit{RBS} elemental distribution line profile along PdHT cross section.}
\label{fig:mechanism}
\vspace{-0em}%
\end{figure*}

We proposed a possible explanation for the forming-free behavior (see Fig.~\ref{fig:mechanism}a) based on a suite of comprehensive characterizations, including the Pd/HfO\textsubscript{2-x} interface imaging by \textit{HRSTEM-iDPC} (\textit{iDPC} stands for integrated Differential Phase Contrast), which allows us to image light atoms in the presence of heavy atoms (as shown in Fig.~\ref{fig:mechanism}b and Fig.~\ref{esi_O_ions}a), the surface profiles of Pd and Pt thin films obtained via Atomic Force Microscopy (\textit{AFM}) (Fig.~\ref{esi_afm_dev}a, b), the Arrhenius activation energy measurements for both PdHT and PtHT memristors using a semiconductor analyzer (Fig.~\ref{fig:mechanism}c), and the elemental distribution profiles acquired from Rutherford Backscattering Spectroscopy (\textit{RBS}) (Fig.~\ref{fig:mechanism}d). 

The activation energies for the pristine PtHT and PdHT devices are compared in Fig.~\ref{fig:mechanism}c \textit{via} the Arrhenius equation, $\rho(T) = \rho_0 \exp\left({E_a}/{k_B T}\right)$, where \( \rho(T) \) denotes the resistivity, \( T \) is the absolute temperature in Kelvin, \( \rho_0 \) is the pre-exponential factor, \( E_a \) represents the activation energy, and \( k_B \) is the Boltzmann constant. As illustrated in Fig.~\ref{fig:mechanism}c, the activation energy exhibits two distinct regimes: a Low-Temperature (\textit{LT}) regime (300 K to 400 K) marked by lower activation energies (denoted as \(E_{a1}\)) and a High-Temperature (\textit{HT}) regime (400 K to 470 K) characterized by larger activation energies (denoted as \(E_{a2}\)). 

Notably, the pristine PdHT devices exhibit activation energies of \(E_{a1} = \SI{0.055(8)}{eV}\) and \(E_{a2} = \SI{0.22(2)}{eV}\), which are significantly lower than those of the PtHT devices (\(E_{a1} = \SI{0.124(4)}{eV}\) and \(E_{a2} = \SI{0.502(6)}{eV}\)). 
The activation energy \(E_{a2}\) of PtHT is possibly attributed to the diffusion barrier associated with the migration of doubly positively charged oxygen vacancies inside HfO\textsubscript{2-x}, in accordance with previous reports~\cite{capron2007migration,traore2015hfo,mueller2018sims,clima2012first}. It may arise from the diffusion barrier of doubly negatively charged interstitial oxygen ions, based on existing \textit{ab initio} calculations~\cite{hou2008induced,traore2015hfo,clima2012first}. Meanwhile, it is noteworthy that the Frenkel-Pair (\textit{FP}) dissociation process in HfO\textsubscript{2} typically requires activation energies on the order of \SI{5.2}{eV}~\cite{traore2015hfo, ottking2015defect}. Furthermore, if the \textit{FP} process comprises a doubly positively charged pair (\(V_{\mathrm{O}}^{2+}\) + \(O^{2-}\)), its formation energy is \SI{5.8}{eV}~\cite{alex2002vacancy}, a value closely related to the electroforming process, as reported previously~\cite{traore2015hfo, waser2007nanoionics}, implying that \textit{FP} dissociation is highly unlikely to occur under both \textit{LT} and \textit{HT} conditions without an applied voltage bias. 

In pristine PtHT devices, the presence of \(V_{\mathrm{O}}^{2+}\) is expected in amorphous hafnia devices and is introduced during the sputtering process~\cite{calka2014engineering}, though it remains at a low level (see more details in the ESI part 1 about doubly positively charged oxygen vacancies formation). In contrast, in PdHT devices, activation energy appears to be attributed to singly positively charged (\(V_{\mathrm{O}}^{+}\)) defects or interstitial oxygen \(O_i^{2-}\) defects. This difference is likely due to the incorporation of Pd in HfO\textsubscript{2-x} at the Pd/HfO\textsubscript{2-x} interface (as observed in Fig.~\ref{fig:mechanism}b), adsorbing the oxygen ions around the Pd atoms (Pd–\(O_i^{2-}\)), generating more oxygen vacancies along the conductive path. Interstitial Pd atoms alter the defect landscape in hafnia by creating more oxygen vacancies and donating electrons, thereby stabilizing oxygen vacancies as singly positively charged (\(V_{\mathrm{O}}^{+}\)) defects and forming \(V_{\mathrm{O}}^{+}\)–\(O_i^{2-}\) pairs, which serve as traps for electrons hopping~\cite{alex2002vacancy}. Additional activation energy measurements on the \textit{HRS} state of both PdHT and PtHT memristors (Fig.~\ref{esi_ea}) demonstrate that after electroforming (i.e., completion of \textit{FP} dissociation) in PtHT devices, the dominant charges exhibit an activation energy of approximately \SI{0.22(2)}{eV}, similar to that observed in PdHT devices, which is also consistent with the fact that after electroforming of the PtHT devices, the IV characteristics of both types of devices are basically indistinguishable.


Considering above all, the forming-free phenomenon likely originates from Pd atoms initially forming a conductive bridge through bonding oxygen ions from hafnia, generating more oxygen vacancies (see Fig.~\ref{fig:mechanism}a). The incorporation of Pd supplies electrons that effectively reconstruct oxygen vacancies/ions in the Pd–O–Hf configuration relative to the conventional Hf–O configuration in PtHT memristors. Furthermore, under minimum bias, it provides the maximum electrons among the Pd bridges, and its inherent special configuration provides electron injection by effectively providing electrons to defective sites and eases the forming process~\cite{traore2015hfo} and conductive path for the initial states. 


To substantiate the hypothesis of the Pd-induced conductive bridges in HfO\textsubscript{2-x} of PdHT, \textit{RBS} measurements were conducted. As depicted in Fig.~\ref{fig:mechanism}d, the HfO\textsubscript{2-x} layer in the PdHT memristors exhibits a higher Pd concentration compared to the Pt concentration in PtHT devices. Remarkably, even after annealing the PdHT and PtHT devices at 573 K for 5 minutes, the Pd atomic ratio within the HfO\textsubscript{2-x} layer remains higher than that of Pt in the PtHT devices, as further illustrated in Fig.~\ref{esi_rbs_eels}a. This observation agrees with the lower migration barrier of Pd into semiconductors compared with other metals such as Li, Cu, Ag, Pt, and Au~\cite{tahini2015ultrafast}.

We employed \textit{HRSTEM-HAADF} to examine the Pd/HfO\textsubscript{2-x} interface. As shown in Fig.~\ref{fig:mechanism}b, the images reveal that the Pd-Hf atoms are clearly observed above Pd electrodes taken in the [011] zone, showing a region where crystallized HfO\textsubscript{2-x} (evidenced by small, moderately bright atoms, marked in purple spheres as shown in the most right \textit{HRSTEM-iDPC} image) intermixed with Pd (characterized by larger, highly bright atoms, marked in orange spheres), which is further emphasized in Fig.~\ref{esi_O_ions}b with red cycles. For further validation of the Pd–O–Hf configuration, please refer to ESI part 2 for further discussion on it.

The influence of Pd on forming voltages was examined via the I–V characteristics of various devices with various electrodes (see Figs.~\ref{esi_electrodes}a, b). The critical role of Pd electrodes is underscored by their ability to markedly reduce or even eliminate the need for electroforming voltages, in stark contrast to the PtHT structure (Fig.~\ref{esi_electrodes}c). Moreover, as the HfO\textsubscript{2} thickness increases, a distinct electroforming behavior emerges (Fig.~\ref{esi_electrodes}d), \begin{figure*}[!t]
\centerline{\includegraphics[width=1.0\linewidth]{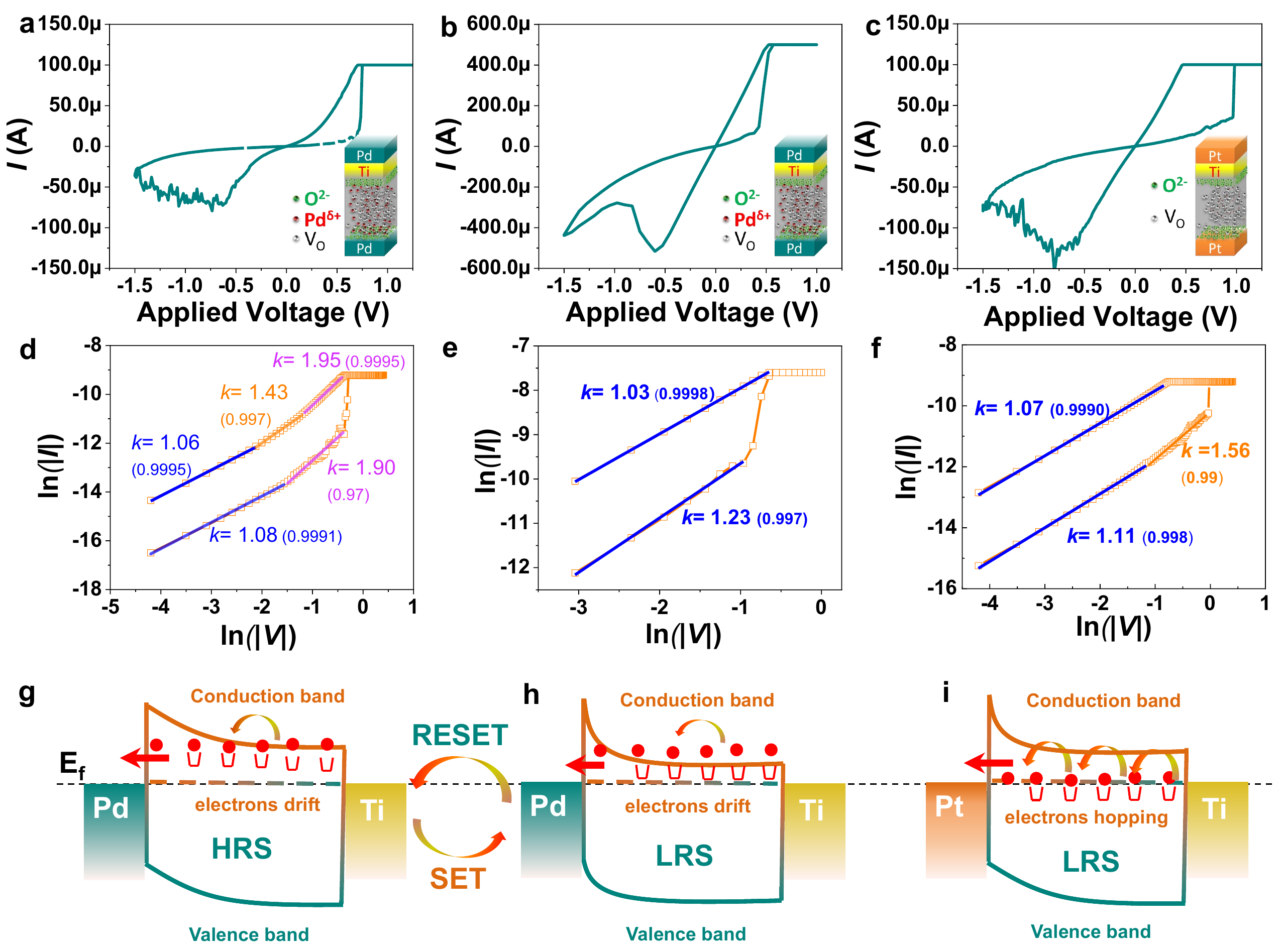}}
\vspace{-0.0em}%
\caption{\textbar\quad \textbf{Investigation of conduction mechanism from I-V curves fitting for PtHT and PdHT.} \textbf{a} I-V curves of PdHT device with 0.1 mA compliance current, \textbf{b} 0.5 mA compliance current.
\textbf{c} Present PtHT device I-V curve with 0.1 mA compliance current. \textbf{d, e, f}  The fitted I-V plots of positive bias parts \textbf{a, b, c}, respectively. \textbf{g, h} Illustration of \textit{HRS} and \textit{LRS} band diagrams of PdHT and electrons drifting between shallow vacancy states, respectively. \textbf{i} The band diagram of PtHT \textit{LRS} states and electron hopping between deep defective sites.}
\label{fig:sclc}
\vspace{-0.0em}%
\end{figure*}likely due to the finite extent of the Pd–O–Hf configuration within HfO\textsubscript{2}, which results in an exponential increase in resistance with a linear increase in thickness.


\subsubsection*{Investigation of conduction mechanism}


In \textit{ReRAM} devices, multiple conduction mechanisms~\cite{wang2018conduction,funck2021comprehensive}, such as Ohmic conduction, trap-mediated conduction, Trap-Assisted Tunneling (TAT), Space-Charge-Limited Conduction (\textit{SCLC}), Poole–Frenkel emission, Schottky emission, and valence change-based mechanisms, collectively dictate the overall device behavior.
In particular, \textit{SCLC} is a prominent mechanism observed in \textit{ReRAM} devices during high-voltage regime~\cite{funck2021comprehensive}. Here, the current is constrained by a space-charge-limited region in which charge carriers, injected from the electrodes, traverse localized trap states within the resistive switching material. The \textit{SCLC} behavior is critically influenced by the trap density and trap energy levels, which govern the mobility and transport of the charge carriers.

A trap-controlled \textit{SCLC} regime can be segmented into two distinct portions: the trap-unfilled (trap-limited) regime and the trap-filled (trap-free) regime. In the Mott-Gurney region, characterized by the power law \(I\propto V^2\), a steep increase in current is observed at high electric fields. This behavior is typically identified by an initial Ohmic conduction at low fields, followed by a transition to a power-law dependence as electrode-injected electrons surpass the equilibrium concentration. Consequently, \textit{SCLC} conduction is more likely when the electrode contact exhibits high carrier injection efficiency~\cite{funck2021comprehensive}. The \textit{SCLC} conduction mechanism is quantitatively described by the Mott-Gurney law (Eq.~\ref{eq:mott}), which relates the current density \(J\) to the applied voltage \(V\) and material properties:

\begin{equation}
 \fontsize{9pt}{9pt}\selectfont
 J = \frac{4}{9} \epsilon_0 \mu \frac{V^2}{d^3}
 \label{eq:mott}
\end{equation}
Here, \(\epsilon_0\) denotes the dielectric permittivity, \(\mu\) is the carrier mobility, \(d\) represents the film thickness, and \(V\) is the applied voltage.

The slope (\(k\)) of the double logarithmic I–V curve evolves from approximately 1.0 (indicative of trap-limited Ohmic conduction) through 1.5 (reflecting the trap-unfilled transitional region) to nearly 2.0 (characteristic of trap-free \textit{SCLC}) for both \textit{HRS}and \textit{LRS}, as observed in Figs.~\ref{fig:sclc}a and d for PdHT devices under a 0.1 mA compliance current. On the contrary, when the compliance current is increased to 0.5 mA, the slope \(k\) for \textit{HRS} (see Figs.~\ref{fig:sclc}b and e) is dominated by Ohmic conduction in both \textit{HRS} and \textit{LRS}, 
\begin{figure*}[!t]
\centerline{\includegraphics[width=1.0\linewidth]{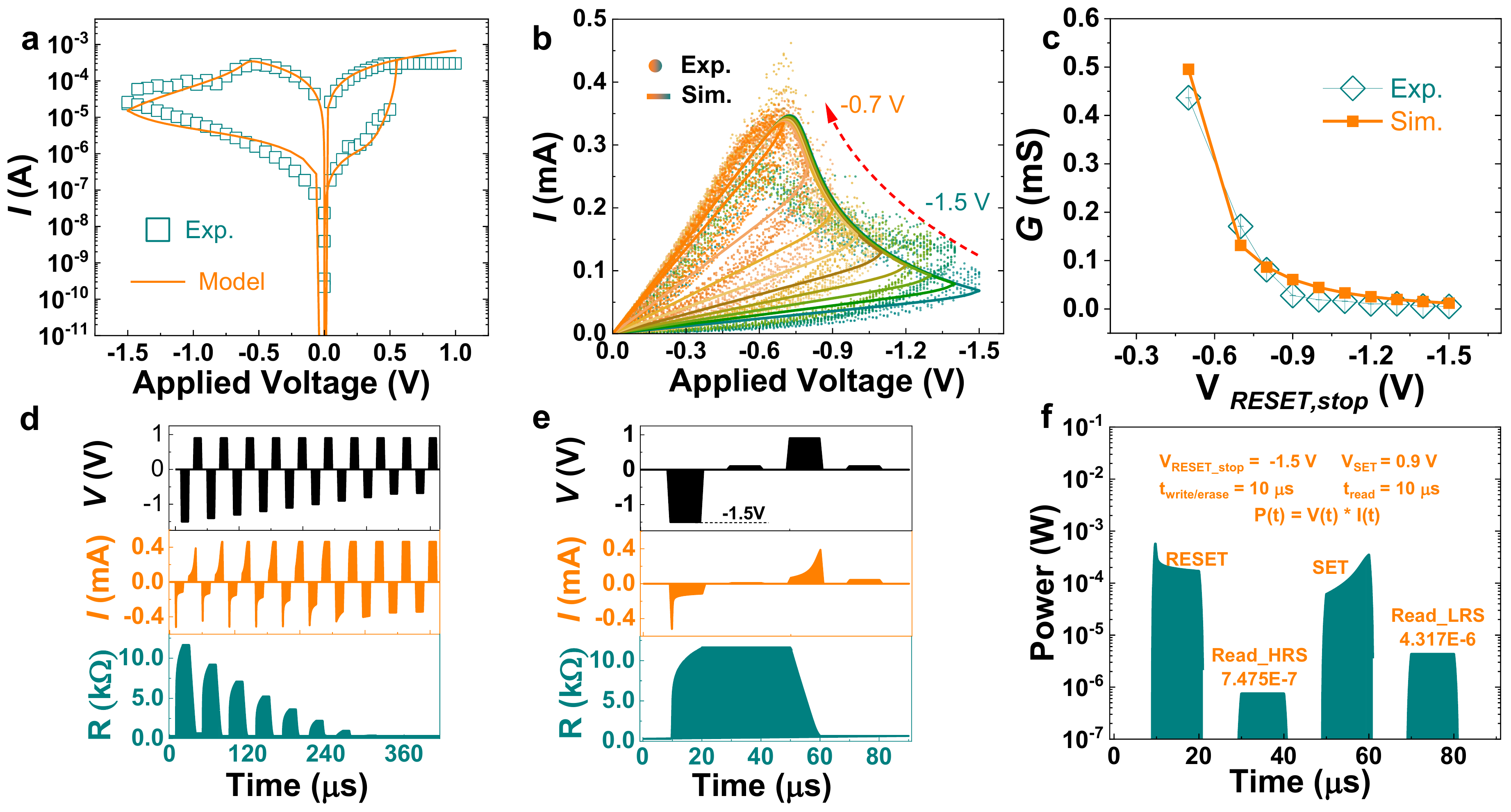}}
\vspace{-0.0em}%
\caption{\textbar\quad \textbf{Power consumption evaluation} \textbf{a} Experimental \textit{DC} I-V curve calibration with \textit{JART} memristor model. \textbf{b} \textit{MLR} verification using \textit{DC} sweeping by changing V\textsubscript{\textit{RESET}}. \textbf{c} Comparison of conductance gradual change for experimental and simulation results. \textbf{d} 8 resistance states programming schemes, including the results of currents and resistance over time. \textbf{e} Programming scheme of the resistance state with V\textsubscript{\textit{RESET, stop}} = \SI{-1.5}{\volt}. \textbf{f} Power consumption evaluation for programming  the conductance states in \textbf{e}.}
\label{fig:model}
\vspace{-0.0em}%
\end{figure*}
a consequence of reduced interface state density at the Pd/HfO\textsubscript{2-x} interfacial region owing to filament formation. Conversely, PtHT devices exhibit ohmic conduction under a 0.1 mA compliance current for both \textit{HRS} and \textit{LRS} (Figs.~\ref{fig:sclc}c and f).These results indicate that at the Pd/HfO\textsubscript{2-x} interface in PdHT devices, shallow defect states are preferentially formed, enabling electrons to enter the conduction band (see Figs.~\ref{fig:sclc}g and h). In contrast, PtHT devices tend to form deep defect states, wherein electrons tunnel into vacancy defect sites~\cite{funck2021comprehensive} (see Fig.~\ref{fig:sclc}i). 

Taking all the aforementioned analysis in Fig.~\ref{fig:mechanism} and Fig.~\ref{fig:sclc}, including the activation energy study, \textit{HRSTEM} observations of Pd atoms, \textit{EELS}and \textit{RBS} elemental distribution profiles of PdHT, and I-V curve fitting, into consideration, in PdHT devices, the Pd–O–Hf configuration, including the Pd-induced conductive bridges and electron injection from Pd atoms in HfO\textsubscript{2-x}, facilitates oxygen vacancies reconfiguration and electron drift toward the Pd electrode via tunneling. It hereby contributes to the elimination of the electroforming process. After the RESET operation for PdHT devices (from initial state to HRS), the Pd-induced conductive bridges will be partially damaged irreversibly by scavenging the oxygen vacancies through oxygen ions from TiO\textsubscript{x}, leaving the shallow defective states thanks to the oxygen affinity nature of Pd. In contrast, in PtHT devices, electrons predominantly hop between deep defect sites generated by high oxygen vacancy concentrations after the electroforming process under higher electric fields.

\subsubsection*{Device calibration and electrical simulation}

To evaluate the energy consumption of the PdHT and PtHT devices, we calibrated the \textit{DC} I–V curves of the real devices using the \textit{JART} model~\cite{jart_model2020}. In the case of PdHT as illustrated in Fig.~\ref{fig:model}, the simulated I–V performance exhibited excellent concordance with the experimental results (see Fig.~\ref{fig:model}a). Regarding the multilevel resistance state performance, the rupture of the \textit{CF} displayed a well-controlled, gradual transition upon tuning the V\textsubscript{\textit{RESET, stop}}, as demonstrated in Fig.~\ref{fig:model}b, where multiple distinguishable I–V curves were observed from Cadence simulation, which implies that the results of PdHT device calibration agree well with experimental results. Similarly, the incremental change in conductance by varying bias amplitude for V\textsubscript{\textit{RESET, stop}} showed remarkable agreement between experimental observations and simulation results (Fig.~\ref{fig:model}c). Furthermore, distinct resistance states can be programmed using variable \textit{RESET} stop voltages with a 10~\(\mu\)s pulse duration, leading to the conclusion that 8 discernible resistance states can be established with corresponding programming schemes, as depicted in Fig.~\ref{fig:model}d. The detailed programming protocol was further illustrated in Fig.~\ref{fig:model}e, where each programmed resistance state is successfully achieved using a writing/reading pulse of 10~\(\mu\)s at $-1.5~V$/100~mV, respectively. To assess the power consumption during programming, we integrated the instantaneous power \(P(t) = I(t)\cdot V(t)\) over time, as shown in Fig.~\ref{fig:model}f. Here, both \textit{RESET/SET} energy consumptions contribute to the total programming energy, while the sum of \textit{HRS} and \textit{LRS} reading energies represents the total energy consumption during reading. The programming energy consumption for 8 distinct conductance states for both PdHT and PtHT devices is summarized in Table~\ref{tab:pdht-energy} and Table~\ref{tab:ptht_energy}, respectively.

\subsubsection*{System-level energy simulation for neuromorphic applications}

\begin{figure*}[!t]
\centerline{\includegraphics[width=1.0\linewidth]{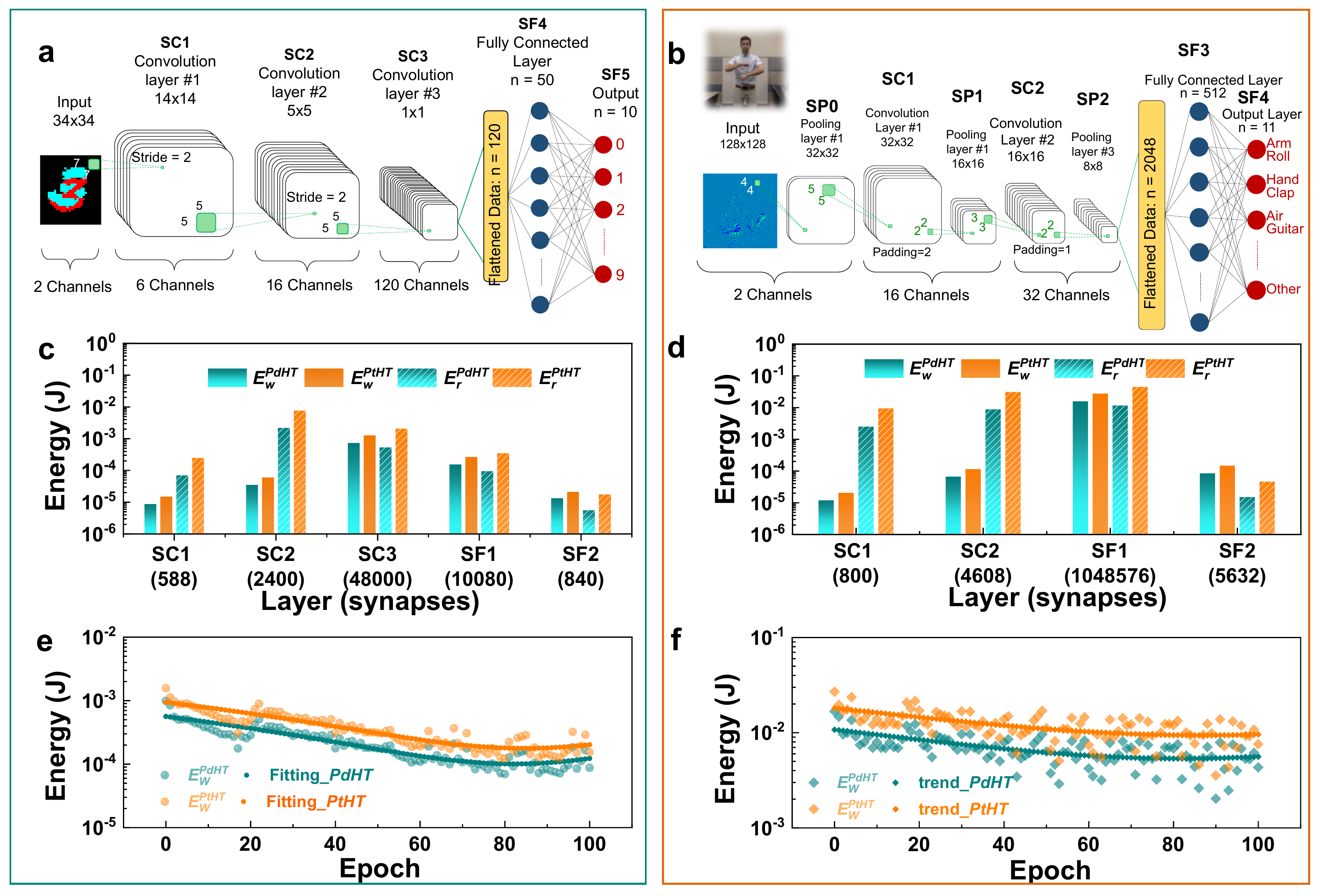}}
\vspace{-0.0em}%
\caption{\textbar\quad \textbf{System-level energy simulation for neuromorphic applications} \textbf{a, b} 
Model architectures of the \textit{N-MNIST} and Gesture \textit{SNNs}, respectively. \textbf{c, d} Overall per-layer write energy consumption after one-time programming of pre-trained \textit{SNN} instances and read energy consumption during the inference of the test set for the \textit{N-MNIST} and Gesture \textit{SNNs}, respectively. \textbf{e, f} Overall write energy consumption per training epoch in the case of online learning for the \textit{N-MNIST} and Gesture \textit{SNNs}, respectively.
}
\label{fig:snn}
\vspace{-0.0em}%
\end{figure*}

For the demonstration of the proposed PdHT \textit{ReRAM} device in realistic neuromorphic applications, we designed two deep convolutional \textit{SNNs} targeted at the classification of the \textit{N-MNIST}~\cite{OJCT15} (Fig.~\ref{fig:snn}a) and IBM \textit{DVS128 }Gesture~\cite{ATBM17} (Fig.~\ref{fig:snn}b) datasets, respectively. These networks are implemented in Python using primitives from the open-source Spike LAYer Error Reassignment (\textit{SLAYER})~\cite{shor18} and PyTorch~\cite{Pytorch19} frameworks, and they are trained with a variant of the backpropagation algorithm. Spiking neurons are modeled on the Spike Response Model (\textit{SRM}), an advanced generalization of the ubiquitous Integrate-and-Fire (\textit{I\&F}) neuron model~\cite{Ge95}. At the output, the class corresponding to the neuron that produced the highest number of spikes is selected as the winning class. For further methodological details, please refer to the Methods section.

To estimate the energy consumption at the system level, we perform a series of experiments and report the overall read and write energy measured collectively for all \textit{ReRAM} devices needed to store the synaptic weights across each network. In the first set of experiments, we train the two \textit{SNNs} in software, and then we write the final synaptic weights directly onto the \textit{ReRAM} devices. For simplicity and generalization purposes, there is no limit to the maximum number of employed devices. The first two bars in Figs.~\ref{fig:snn}c and \ref{fig:snn}d report the write energy, $E_{write}$, per layer on a logarithmic scale for the \textit{N-MNIST} and Gesture \textit{SNNs}, respectively. As anticipated, the write energy scales proportionally with the number of synapses, since a greater number of weights necessitates more devices. Notably, the PdHT devices achieve an overall reduction in write energy of approximately 43\% with respect to that of PtHT for each \textit{SNN}.

In the case of online learning, the synaptic weights are updated and written back on the \textit{ReRAM} devices at every epoch of the training of each \textit{SNN} and the overall write energy per epoch is illustrated in Figs.~\ref{fig:snn}e and \ref{fig:snn}f for the two case studies, respectively. Initially, the synaptic weights are set to small random values, resulting in a peak write energy equal to the total write energy occurring by summing the corresponding bars for all layers in Figs.~\ref{fig:snn}c and \ref{fig:snn}d. This is because in the first iteration almost all \textit{ReRAM} devices need to be programmed to a new value. As training progresses, only those devices that require updates contribute to the write energy, leading to a gradual reduction that mirrors the convergence of the synaptic weights, thereby lowering the overall energy consumption.

Regarding the read energy~$E_{read}$, during inference, represented by the rightmost two bars in Figs.~\ref{fig:snn}c and \ref{fig:snn}d, the energy consumption per layer exhibits a pattern distinct from that of the write energy. In \textit{SNNs}, each layer functions as a filter that reduces the number of spikes transmitted to subsequent layers, resulting in progressively sparser spiking activity. Therefore, the read energy depends not only on the number of synapses but also on the density of incoming spikes. For example, layer SF1 of the Gesture \textit{SNN} consumes only 4 $\times$ more energy than its predecessor, despite having 228 $\times$ more synapses. Overall, both networks exhibit an approximate 73\% reduction in read energy during the complete inference of their respective test sets. The average energy per spike during inference is summarized in Table~\ref{tab:read_energy_spike} and demonstrates a comparable energy reduction when utilizing the proposed PdHT devices over PtHT ones.


\section*{Discussion}\label{sec3}


In this work, we reported the successful \textit{CMOS}-compatible fabrication of PdHT \textit{ReRAM} devices, PdNeuRAM. High-resolution imaging analyses, including \textit{SEM}, \textit{HRSTEM}, and \textit{AFM}, confirm well-defined device stacks with minimal interfacial roughness. Electrical characterization reveals \emph{forming-free} switching at relatively low operating voltages, with \(\mathrm{V_{SET}}\) around 0.56\,V and \(\mathrm{V_{RESET}}\) averaged at \(-0.58\)\,V . Notably, the PdHT devices exhibit reduced \textit{C2C} and \textit{D2D} variability comparable to their PtHT counterparts, indicating robust switching behavior and enhanced reliability.


From previous studies on oxygen chemisorption on Pd and Pt films and Schottky barriers ~\cite{oxidase_on_Pd2015} and \textit{MIGS}~\cite{sbh2007control,migs2000}, it seems unlikely that the Schottky barrier differences cause the electroforming differences. The absence of an electroforming step in PdHT devices is primarily attributed to the presence of Pd atoms within the HfO\textsubscript{2-x} layer for a few nanometers, which probably contributes to the charge redistributions, acting as Pd-induced conductive bridges, thus eliminating the need for a high-voltage electroforming process. Furthermore, regardless of the role of Pd in the conduction mechanism, the PdHT presents similar resistive switching properties to that of PtHT as indicated by the identical activation energy, which is in turn in agreement with the behavior reported in the previous HfO\textsubscript{2-x} devices~\cite{wang2022forming,chen2010ultrathin}.

Current-voltage (\(I\)-\(V\)) analysis indicates that both PdHT and PtHT devices operate under a \textit{SCLC} mechanism. At low bias, conduction follows an ohmic behavior. As the voltage increases, trap-mediated processes dominate; injected carriers fill shallow defect states, transitioning the conduction to a quadratic dependence on voltage (\(I \propto V^2\)), consistent with the Mott-Gurney law. In PdHT devices, intrinsic insulating layer defects and interfacial imperfections, enhanced by Pd-O-Hf interactions, create shallower defect states that promote efficient electron transport and lower variability. In contrast, PtHT devices exhibit deeper defect states, which vary with electroforming voltage and material modifications, leading to increased variability.

The integration of these forming-free multibit \textit{ReRAM} devices into neuromorphic systems has been successfully demonstrated by employing them as synaptic elements in \textit{SNNs}. Using three-device (3×3-bit) configurations to represent 9-bit quantized synaptic weights, the networks are implemented for real-world pattern recognition tasks on the N-MNIST and IBM’s \textit{DVS128} Gesture datasets. The \textit{SNN} architectures, developed using frameworks such as SLAYER and PyTorch, achieve competitive classification accuracies (94.6\% for N-MNIST and 85.6\% for IBM's \textit{DVS128} Gesture). However, energy consumption analyses reveal that the PdHT devices yield substantial energy savings, with write and read operations reduced by approximately 43\% and 73\%, respectively. This energy efficiency, combined with stable multilevel conductance programming and robust endurance, makes the proposed PdNeuRAM technology highly promising for next-generation, low-power neuromorphic computing applications.


\section*{Methods}\label{sec4}
\addcontentsline{toc}{section}{Methods}

\subsubsection*{\textit{ReRAM} cell fabrication} 

 First, the SiO\textsubscript{2} substrate was sequentially cleaned by fuming nitric acid, acetone, isopropyl alcohol, and deionized (\textit{DI}) water in an ultrasonic oscillator. Then, the metal films were deposited on the chemically cleaned SiO\textsubscript{2} substrate through electron beam evaporation in a vacuum chamber with the pressure of 10\textsuperscript{-8} torr. \SI{5}{\nano\meter} Ti (the adhesion layer) and \SI{50}{\nano\meter} Pd (\textit{BE}) were deposited by \textit{ATC 2400} Sputtering System under a vacuum level of 10\textsuperscript{-8} torr;  \SI{5}{\nano\meter} or \SI{10}{\nano\meter} HfO\textsubscript{2} (the oxide layer) was deposited by Sputter (Alliance Concepts system) under a vacuum level of 10\textsuperscript{-7} torr; \SI{5}{\nano\meter} or \SI{10}{\nano\meter} Ti (interface layer) and 50nm Pd (\textit{TE}), which were deposited by \textit{ATC 2400} Sputtering System under a vacuum level of 10\textsuperscript{-8} torr. By lifting off technique, the nodes with dimensions are $\SI{5}{\micro\meter} \times \SI{5}{\micro\meter}$ are fabricated by negative photoresist (\textit{AZnLof@2020}) and positive photoresist (\textit{S1805}) for photolithography (Heidelberg microMLA) (see the profiles in Fig.~\ref{esi_resist}). The pads dimensions are $\SI{15}{\micro\meter} \times \SI{15}{\micro\meter}$. The \textit{BE} via was etched by CHF\textsubscript{3}/Ar gases (Sentech Etchlab 200) for 270 s. The optical microscopy image of the device was obtained by an optical microscope (Olympus BX51) $50 \times$. Scanning Electron Microscope (Hitachi S-4800)  was applied to observe the details of the nodes and cross-section.

\subsubsection*{Electrical performance characterization}
To study the electrical performance of both structures cells, a probe station (\textit{CASCADE}) equipped with  a semiconductor analyzer (Keysight \textit{B1500A})  was utilized to conduct the electrical measurement for the \textit{ReRAM} cell at room temperature (\SI{300}{\kelvin}), during which a bias voltage was applied to the \textit{TE} while the \textit{BE} was grounded. The double linear sweeping bias is from 0 V to 1.5 V with a step \SI{100}{\milli\volt} for the SET sweeping process and \SI{0}{\volt} to \SI{-1.5}{\volt} step \SI{-100}{\milli\volt}  for \textit{RESET} sweeping process.

\subsubsection*{\textit{HRSTEM} characterization}

In our study, we implemented a multifaceted STEM methodology combining \textit{HAADF}, \textit{EELS}, and \textit{iDPC} techniques to achieve a comprehensive characterization of advanced materials. We began with \textit{STEM-HAADF} imaging, which utilizes a high-angle annular dark-field detector to collect thermally diffuse scattered electrons. This method provides Z-contrast images where the intensity scales roughly as Z\textsuperscript{1.7}, allowing us to resolve atomic columns and differentiate heavy and light elements based on their scattering power. Precise alignment and calibration are performed with dwell times in the order of \(\SI{}{\micro\second}\) to achieve high resolution and reliable compositional mapping.

\textit{STEM-EELS} was employed to complement the HAADF images by providing elemental and chemical state information through electron energy loss spectroscopy. \textit{EELS} spectra, collected in parallel with HAADF imaging, enabled the identification of subtle variations in bonding and composition at the nanoscale.

To enhance the detection of light elements and obtain electrostatic potential maps, we applied the  \textit{iDPC} method. In this mode, the segmented DF4 detector is selected via the Velox \textit{DPC/iDPC} control panel, and the resulting differential signals are integrated. This process improves image contrast and reduces sensitivity to defocus and thickness variations, yielding an image where contrast is roughly proportional to atomic number.


\subsubsection*{\textit{RBS} characterization}

 \textit{RBS} was employed to determine the elemental depth profiles of the layers
in the forming-free PdHT devices. \textit{RBS} measurements were conducted using a \textit{Kobe Steel HRBS-V500}
system with a He\textsuperscript{+} ion beam accelerated at 400\,keV. The beam was directed onto the sample at an
incidence angle of 45°, and backscattered ions were detected at a scattering angle of 107.5°. A 512
channel detector with an energy resolution of 2\,keV was used. The \textit{RBS} data was analyzed using the Kobe Steel AnalysisIB software. The measurements revealed the Pd ratio in the HfO\textsubscript{2-x} layer,
confirming the intermixing of Pd-O-Hf. This configuration is critical for reducing the oxygen diffusion
barrier and facilitating the forming-free behavior.

\subsubsection*{Activation energy measurement}

The activation energy of the forming-free HfO\textsubscript{2}-based \textit{ReRAM} devices was measured using a Keithley \textit{B1500A} Semiconductor Parameter Analyzer in conjunction with a thermionic heater. In this setup, the devices were mounted on a temperature-controlled stage that provided stable and uniform heating over a defined temperature range. To prevent oxidation and moisture interference during high-temperature measurements, a continuous flow of high-purity N\textsubscript{2} gas was maintained throughout the experiment. The \textit{B1500A} recorded the current-voltage (I-V) characteristics at incremental temperature steps. The temperature-dependent resistivity was then derived from the I-V data, and an Arrhenius plot of $\ln(\rho)$ versus $1/T$ was constructed. The slope of the linear fit, when divided by Boltzmann's constant ($k_B$), yielded the activation energy. This method ensured precise thermal control and reproducible measurements, critical for understanding the conduction mechanisms in the devices.


\subsubsection*{Device calibration and power consumption evaluation }

The device model is developed based on the JART model, which accurately replicates the electrical behavior of the forming-free PdHT \textit{ReRAM} devices. Calibration of the model is achieved by fitting the experimental \textit{DC} I–V curves, obtained using a Keisight \textit{B1500A}, with the simulation results. In our setup, both \textit{DC} sweep and pulse-based measurement configurations are implemented using Cadence simulation tools. The \textit{DC} sweep methodology enables the extraction of key parameters such as \textit{SET/RESET} voltages and memory window, while pulse-based simulations reveal the dynamic switching characteristics and multi-resistance states under various programming schemes. This integrated simulation approach provides critical insights into the energy consumption during programming and reading operations, ensuring consistency between experimental and simulated results.

\subsubsection*{Device-based system-level \textit{SNN} simulation}

The synaptic weights of the two \textit{SNNs} are quantized to integer values of 9 bits, which are represented by a combination of three multistate \textit{ReRAM} devices with 8 states each, i.e., 3$\times$3$-$bit \textit{ReRAM} devices. In the presented experiments, we performed the training and inference for the two case studies and reported the write and read energy footprint of the PdHT and PtHT \textit{ReRAM} devices in different scenarios. The focus is kept on the energy impact of the crossbar arrays hosting the devices since it is not dependent on the actual realization of the periphery circuitry, allowing space for generalization no matter the actual implementation of the neuromorphic hardware accelerator. In all cases, we present a comparison between state-of-the-art PtHT \textit{ReRAM} devices and the proposed PdHT ones.

\textbf{\textit{N-MNIST SNN:}}
The \textit{N-MNIST} dataset is a neuromorphic, that is, spiking, version of the MNIST dataset, which comprises images of handwritten arithmetic digits in grayscale format~\cite{OJCT15}. It consists of 7$\times$10$^4$ sample images that are generated from the saccadic motion of a Dynamic Vision Sensor (\textit{DVS}) in front of the original images in the MNIST dataset. The samples in the \textit{N-MNIST} dataset are not static, i.e. they have a duration in time of 300 ms for each. The dataset is divided into a training set of 6.0$\times$10$^4$ samples and a test set of 1.0$\times$10$^4$ samples. The \textit{SNN} architecture shown in Fig.~\ref{fig:snn}a comprises 3 convolutional layers (SC1, SC2, and SC3) followed by 2 fully connected ones (SF4 and SF5). The classification accuracy on the test set is 94.56\%, which is comparable to the performance of state-of-the-art level-based DNNs.
  
\textbf{Gesture \textit{SNN}:}
The IBM’s \textit{DVS128} Gesture dataset consists of 29 individuals performing 11 hand and arm gestures in front of a \textit{DVS}, such as hand waving and air guitar, under 3 different lighting conditions~\cite{ATBM17}. Samples from the first 23 subjects are used for training, and samples from the last 6 subjects are used for testing. In total, the dataset comprises 1342 samples, each of which lasts about 6s, making the samples 20 times longer than those in \textit{N-MNIST}. To speed up the neuromorphic simulations, we trimmed the length of the samples to about 1.5 s. The proposed \textit{SNN} architecture consists of a pooling layer SP0 to reduce the input samples, 2 convolutional layers (SC1 and SC2) followed by a pooling layer each. The data coming from the last pooling layer is flattened and fed to 2 fully connected layers (SF3 and SF4). The network architecture is presented in Fig.~\ref{fig:snn}b. The network performs with an accuracy of 85.61\% in the test set, which is acceptable considering the shortened samples in the dataset and the shallower architecture compared to the originally proposed architecture in~\cite{ATBM17}.



In both use cases, the \textit{SNNs} are implemented with SRM spiking neurons, effectively capturing the temporal dynamics of spiking activity. Training uses a variant of the backpropagation algorithm, in which error is calculated on the probability of each neuron to change spike state, i.e., fire a spike if it was in a resting state or stop firing, in the next timing instance~\cite{shor18}.




\section*{Data availability}\label{sec5}
\addcontentsline{toc}{section}{Data availability}

The data that support the findings of this study are available from the corresponding author upon reasonable request.


The code that support the findings of this study are available from the corresponding author upon reasonable request.



\begin{appendices}






\end{appendices}


\bibliography{sn-bibliography}

\section*{Acknowledgements}

This work was supported by Delft University of Technology. We acknowledge the kind support of the Delft Kavli Nanofabrication cleanroom, especially Marinus Fischer, Roald van der Kolk, Charles de Boer, and Eugene Straver, for their unconditional assistance. We also specially thank Lennart P. L. Landsmeer and Emmanouil Arapidis for their kind support and discussion.

\section*{Author contributions}
EH, HA, RI:  Conceptualization. EH: device fabrication, electrical measurement. EH, HA, RI, SH, GG: Results Analysis and Visualization. MA, BN, EH, HA: STEM, EELS, and iDPC results acquisition and analysis. EH, BN, HA: AFM, activation energy analysis. EH, HX: Device calibration based on model. TS, EH, HA, AG: SNN simulation, visualization, and analysis. AMS, EH, HA, LB, NE: RBS analysis. EH, BN, TS, and HA: manuscript draft $\&$ finalization. All coauthors contribute to the manuscript review and revision. 

\section*{Competing interests} 

The authors declare no competing interests.
\section*{Additional information}
\subsection*{Supplementary information} 
The online version contains supplementary material available at
https://doi.org/xxx/xxxx.

\newpage


\begin{center}
\section*{\textbf{Electronic Supplementary Information}}
\end{center}

\setcounter{figure}{0}
\renewcommand{\thefigure}{S\arabic{figure}}

\setcounter{table}{0}
\renewcommand{\thetable}{S\arabic{table}}



\begin{figure}[!ht]
\centerline{\includegraphics[width=0.6\linewidth]{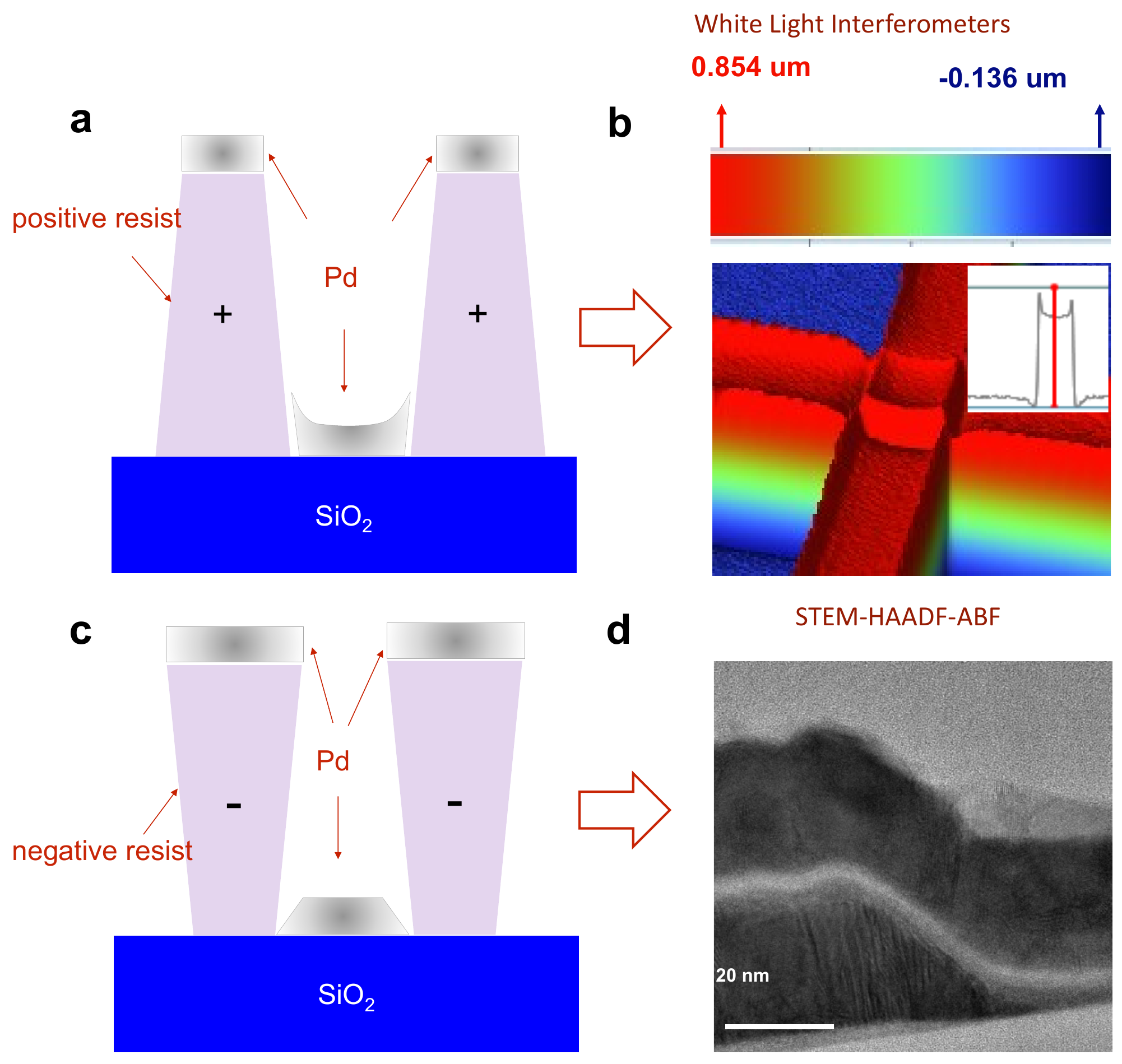}}
\caption{\textbar\quad\textbf{Pd deposition on positive resist (a) and negative resist (c). (b), (d) are the cross section profile of as-fabricated different resists } }
\label{esi_resist}
\vspace{-1em}%
\end{figure}

\begin{figure}[!ht]
\centerline{\includegraphics[width=1.0\linewidth]{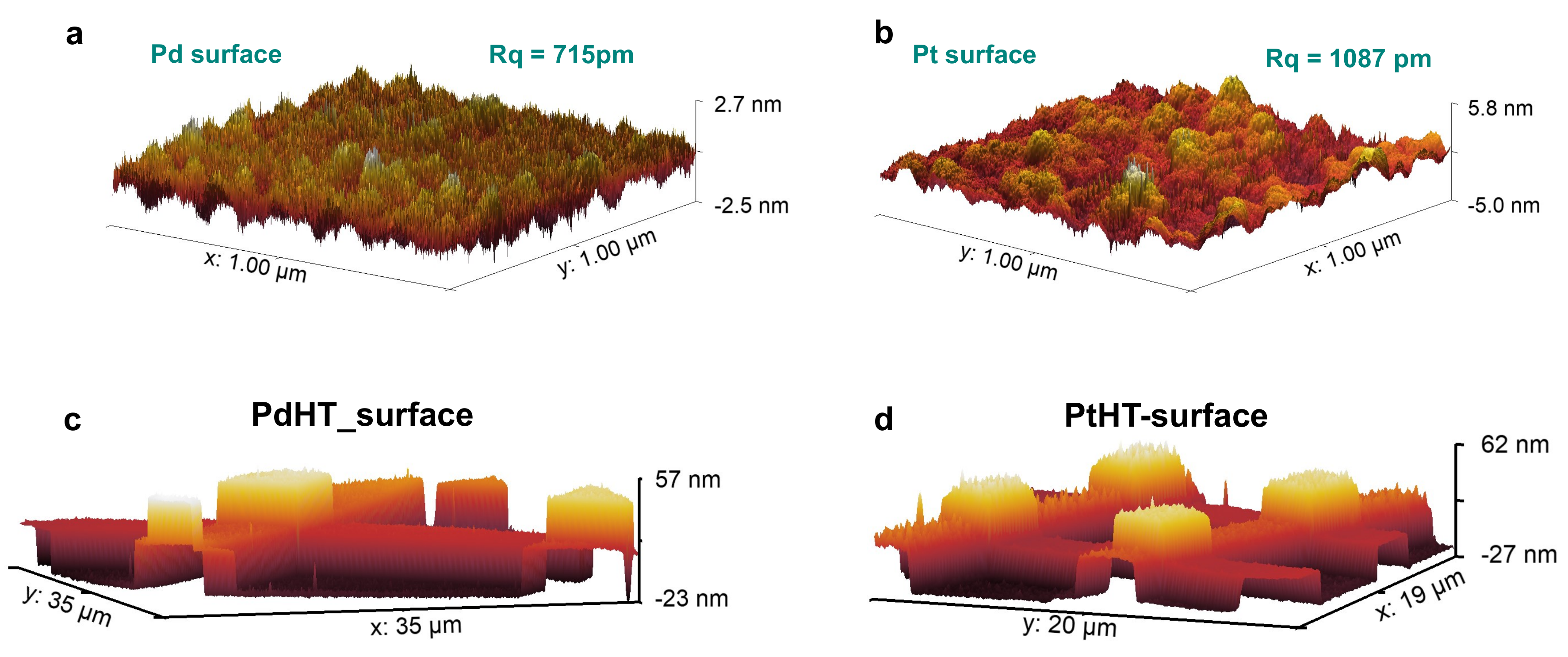}}
\caption{\textbar\quad\  \textbf{a} \textit{AFM} image of Pd surface with a roughness of 715 pm. \textbf{b} \textit{AFM} image of Pt surface with roughness of 1087 pm. \textbf{c} \textit{AFM} surface analysis of PdHT device and \textbf{d} \textit{AFM} surface analysis of PtHT device}
\label{esi_afm_dev}
\vspace{-1em}%
\end{figure}

\begin{figure}[!ht]
\centerline{\includegraphics[width=0.6\linewidth]{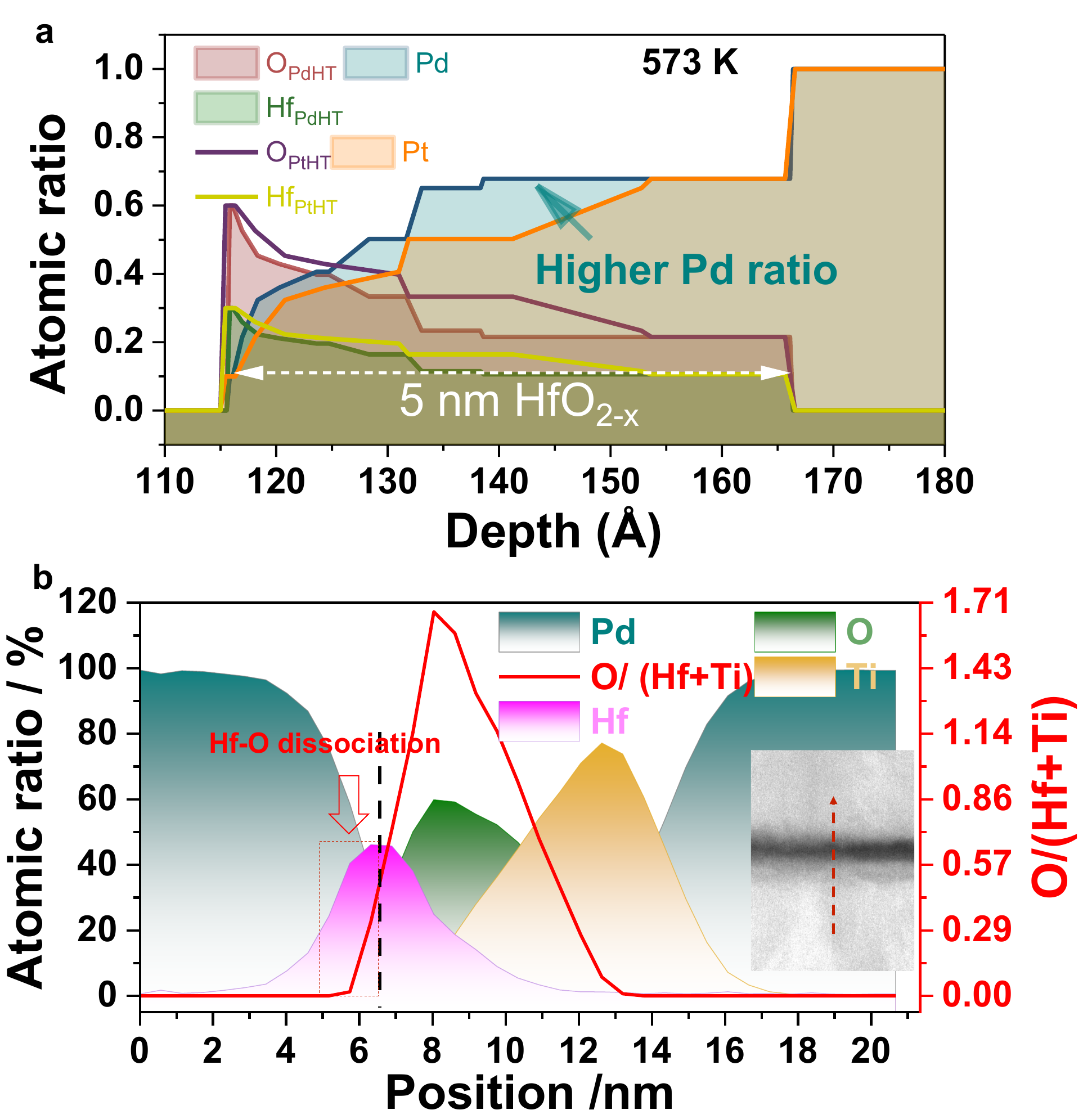}}
\caption{\textbar\quad\textbf{a} Atomic ratio comparison of PtHT and PdHt under 573K (denoted as \textit{573 K}). \textbf{b}  STEM-EELS line profile of \textit{LRS} state for PdHT, where it shows the Hf ions and oxygen ions dissociation, and Hf mixed with Pd layer while oxygen ions migrate into the Ti layer.}
\label{esi_rbs_eels}
\vspace{-1em}%
\end{figure}

\noindent
\begin{minipage}[t]{1.0\textwidth}
    \textbf{1. Doubly positively charged oxygen vacancies formation}
    
    Within pristine PtHT devices, the formation of doubly charged oxygen vacancies, \(V_{\mathrm{O}}^{2+}\), is inherently expected in amorphous hafnia-based PtHT devices. These vacancies are primarily generated during the sputtering process~\cite{calka2014engineering} and through reactive interactions between titanium (Ti) and nonstoichiometric hafnium oxide, HfO\(_{2-x}\). In these reactions, doubly negatively charged oxygen ions combine with Ti to yield the stable electride compound $\mathrm{Ti}_2\mathrm{O}\cdot e^{-}$ at the Ti/HfO\(_{2-x}\) interface, thereby functioning as an effective electron reservoir while concomitantly leaving behind \(V_{\mathrm{O}}^{2+}\) vacancies in HfO\textsubscript{2-x}. The preferential stabilization of Ti-based electrides thanks to the higher electron affinity nature of Ti than that of Hf (according to the work function of the elements), as opposed to the inherently unstable electron-localized HfO\(_{2-x}\) variants without the presence of Ti or TiO\textsubscript{x}, further substantiates this mechanism~\cite{traore2015hfo,hua2023negatively}. Conversely, in PdHT devices, the incorporation of palladium (Pd), which possesses a lower electron affinity relative to platinum (Pt), may promote electron donation. This process facilitates the reduction of \(V_{\mathrm{O}}^{2+}\) to \(V_{\mathrm{O}}^{+}\) while simultaneously enhancing the movement of interstitial oxygen ions \(O_i^{2-}\) with the presence of Ti beyond room temperature so that it will facilitate the formation of TiO\textsubscript{x}.

\textbf{2. Further validation of Pd-O-Hf configuration presence}

     Meanwhile, a finding that is corroborated by \textit{AFM} measurements (Fig.~\ref{esi_afm_dev}a, b). Statistical analysis shows that the Pd film exhibits a roughness of 715 pm, markedly lower than the 1087 pm measured for the Pt film, implying that the Pd surface is less prone to form spikes~\cite{hu2022investigation} and thereby reduces potential measurement errors in \textit{RBS} analysis. From \textit{HRSTEM} observation, as shown in Figs.~\ref{esi_resist}c and d, we can also conclude that there are no Pd spikes generated.  Furthermore, the \textit{STEM-EELS} line profiles confirm the anticipated elemental composition in \textit{LRS} (Fig.~\ref{esi_rbs_eels}b), with the distinct elemental layers aligning precisely with the design illustrated in Fig.~\ref{RRAM_fab}d. Notably, the measurements verify the Pd-O-Hf configuration at the Pd/HfO\textsubscript{2-x} (\textit{LRS}) interface and in the HfO\textsubscript{2-x} layer, a possible consequence of the lower diffusion barrier and higher propensity for PdO\textsubscript{x} formation~\cite{tahini2015ultrafast,do2017study} of Pd.

\textbf{3. MIGS model explanation}

     The MIGS model  attributes \textit{SBH} pinning to a finite density of metal Fermi energy, we use the relation:
        \begin{equation}
        \fontsize{9pt}{9pt}\selectfont
        \phi_{Bn} 
        = S \bigl[\phi_{M} - \chi_{\mathrm{HfO}_2}\bigr]
          + (1 - S)\bigl[E_{\mathrm{CNL}} - E_{\mathrm{CBM}}\bigr]
          \label{eq:migs}
        \end{equation}
where $\phi_{Bn}$ is the Schottky barrier height, $\phi_{M}$ represents the effective work function of the metal, $S$ is the pinning factor at the metal/HfO\textsubscript{2-x} interface, $E_{\mathrm{CNL}}$ is the charge neutrality level of the dielectric layer, $\chi_{\mathrm{HfO}_2}$ is the electron affinity of HfO\textsubscript{2-x}, and $E_{\mathrm{CBM}}$ is the conduction band minimum. Based on this model (with $S = 0.52$, see ref. ~\cite{sbh2007control}), we calculate the \textit{SBH} of PtHT and PdHT,
     
\end{minipage}%

\begin{figure}[!ht]
\centerline{\includegraphics[width=0.8\linewidth]{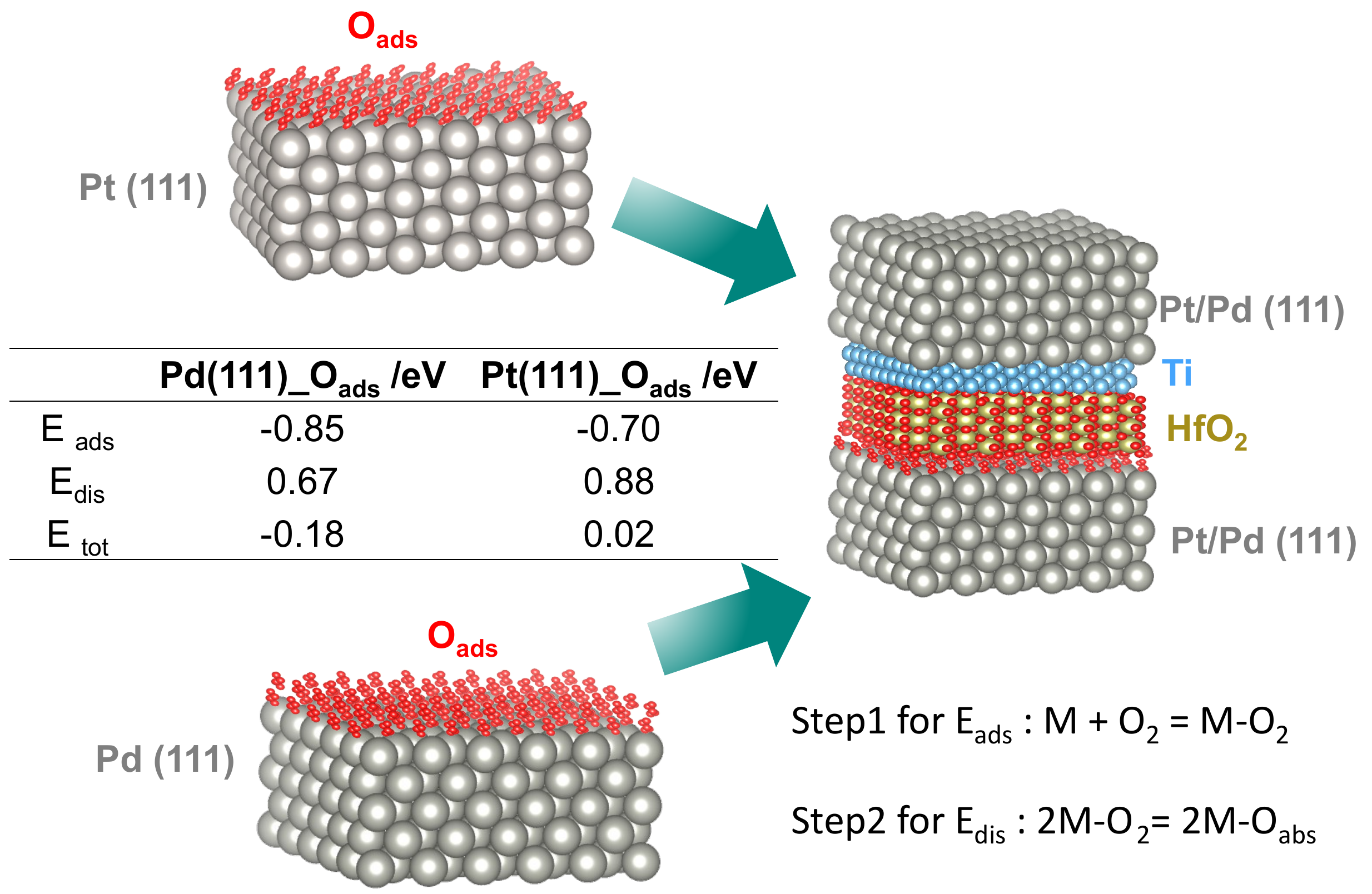}}
\caption{\textbar\quad Atomic schematics of oxygen adsorption on Pd or Pt including adsorption energy and dissociation barriers, according to the previous study~\cite{oxidase_on_Pd2015}. it  }
\label{esi_chemiabsob}
\vspace{-1em}%
\end{figure}

\begin{figure}[!ht]
\centerline{\includegraphics[width=0.6\linewidth]{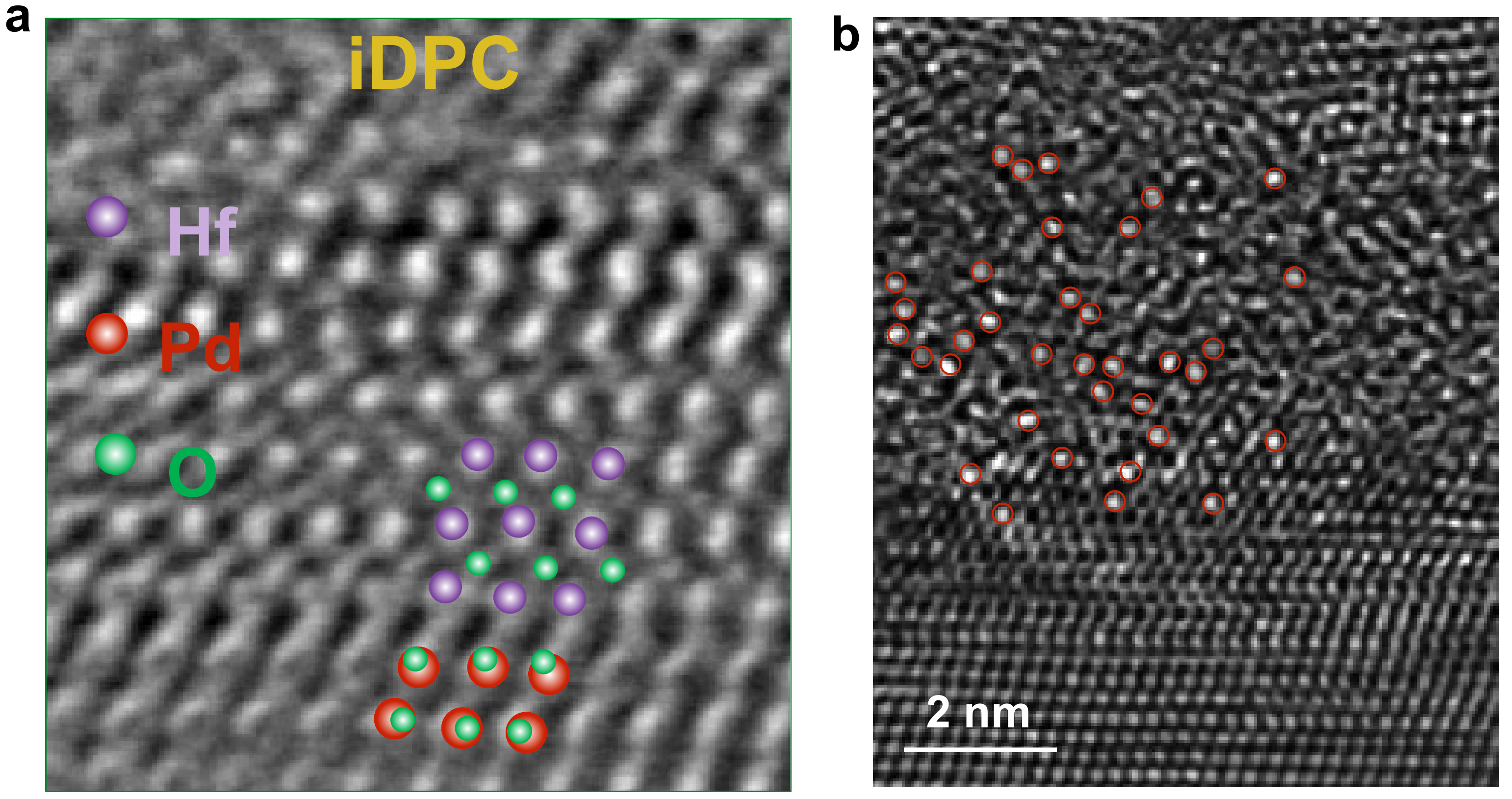}}
\caption{\textbar\quad\textbf{a} Oxygen ions observation using \textit{STEM-iDPC}. \textbf{b} \textit{STEM-iDPC} cross section image of pristine PdHT to indicate the Pd atoms through red cycles across HfO\textsubscript{2-x} layer}
\label{esi_O_ions}
\vspace{-1em}%
\end{figure}

\noindent
\begin{minipage}[t]{1.0\textwidth}

\textbf{4. Study on the SBH for Pd/HfO\textsubscript{2-x} and Pt/HfO\textsubscript{2-x}}
     
     Apart from the aforementioned observations and analyses, we also investigated the band structures and Schottky barriers of the PdHT and PtHT memristors to ascertain whether the forming-free behavior arises from differences in Schottky barrier heights. From previous studies~\cite{yamashita2016situ,giner1953elektronenaustrittsspannungen,jaeckel1963photo,oxidase_on_Pd2015}, it is well established that oxygen can chemisorb on noble metals such as Pt and Pd at room temperature, forming a stable layer of atomic oxygen or surface oxides. This phenomenon occurs due to the strong interaction between oxygen and the d-electrons of these metals under ambient conditions. In Fig.~\ref{esi_chemiabsob}, we summarize the oxygen chemisorption energies on Pd(111) and Pt(111), along with the oxygen dissociation barriers, from which it is evident that Pd(111) exhibits a higher propensity to form Pd(111)-O\textsubscript{abs} than Pt(111). Such behavior significantly influences the work functions of Pd and Pt, as reported in previous studies~\cite{jaeckel1963photo,giner1953elektronenaustrittsspannungen}, where the effective work functions of Pt and Pd are in the ranges of 6.00–6.10 eV and 6.02–6.37 eV, respectively. Moreover, oxygen adsorption affects the work function differently: for Pt, it increases by 0.35–0.45 eV, whereas for Pd, the increase is 0.90–1.25 eV, a disparity attributed to the distinct dipole moments in Pt–O and Pd–O bonding. This distinct impact is corroborated by the observation of oxygen ions via the \textit{STEM-iDPC} technique at the Pd lattice in the Pd/HfO\textsubscript{2-x} interface (Fig.~\ref{esi_O_ions}a). In \textit{HRSTEM-iDPC} images, as shown in Fig.~\ref{esi_O_ions}a, especially when combined with \textit{HAADF} imaging, it serves as a complementary imaging mode. While \textit{HAADF} primarily provides Z-contrast (sensitive to atomic number, thus highlighting heavier elements), \textit{iDPC} collects the differential phase contrast signals, often using a segmented or pixelated detector, and then integrates them to reconstruct the projected electrostatic potential of the sample. This integration allows for enhanced contrast of light elements and detailed structural information that might be missed by \textit{HAADF} alone. From this image, we can observe that at the interface of Pd/HfO\textsubscript{2-x} oxygen is identified among Hf atoms and on Pd atoms.
     
     Furthermore, the Charge Neutrality Level (\textit{CNL}) for HfO\textsubscript{2-x} in both the Pd/HfO\textsubscript{2-x} and Pt/HfO\textsubscript{2-x} systems has been determined experimentally and theoretically, with the extracted \textit{CNL} (E\textsubscript{\textit{CNL}},HfO\textsubscript{2-x}) reported as 4.36 eV~\cite{gu2006effective}. From Metal-Induced Gap States (MIGS)~\cite{sbh2007control,migs2000},  the calculated Schottky barrier heights for Pd/HfO\textsubscript{2-x} and Pt/HfO\textsubscript{2-x} are 4.17–4.22 eV and 4.18–4.37 eV, respectively. This result implies that the difference in activation energy E\textsubscript{a1} observed in Fig.~\ref{fig:mechanism}c between the PdHT and PtHT memristors does not originate from disparities in Schottky barrier height. Consequently, this finding indirectly underscores the critical impact of the electrode materials on the forming-free behavior.

\end{minipage}%


\begin{sidewaystable}
    \small
    \setlength{\tabcolsep}{5.0pt}
    \renewcommand{\arraystretch}{2.0}
    \centering
    \caption{\textbar\quad Energy consumption (PdHT). Measured parameters for various conductance states (G1--G8).}
    \label{tab:pdht-energy}
    \begin{tabular}{
      p{0.9cm}  
      p{1.5cm}  
      C{1.5cm}  
      C{1.5cm}  
      C{1.5cm}    
      C{1.5cm}    
      C{1.5cm}    
      C{1.8cm}    
      C{2.0cm}    
      C{1.3cm}    
      }
    \toprule
      & & & & & & & \multicolumn{2}{c}{\textbf{Averaged Power and Energy}} & \\
    \cmidrule(lr){8-9}
      \textbf{G} & \textbf{R\_states} & 
      \(\mathbf{V_{\text{reset}}}\)(V) & \(\mathbf{V_{\text{read}}}\)(V) & 
      \(\mathbf{P_{\text{read}}}\)(W) & \(\mathbf{t_{\text{read}}}\)(s) & \(\mathbf{E_{\text{read}}}\)(J) &
      \(\mathbf{P_{\text{program}}}\)(W) & \(\mathbf{E_{\text{program}}}\)(J) &
      \(\mathbf{E_{\text{tot}}}\)(J) \\
    \midrule
    G1 & HRS1 & -1.5 & 0.1 & 7.48e-07 & 1.00e-05 & 7.48e-12 & 4.20e-04 & 4.20e-09 & 4.21e-09 \\
    G2 & HRS2 & -1.4 & 0.1 & 9.15e-07 & 1.00e-05 & 9.15e-12 & 3.96e-04 & 3.96e-09 & 3.97e-09 \\
    G3 & HRS3 & -1.3 & 0.1 & 1.14e-07 & 1.00e-05 & 1.14e-12 & 4.80e-04 & 4.80e-09 & 4.80e-09 \\
    G4 & HRS4 & -1.2 & 0.1 & 1.44e-06 & 1.00e-05 & 1.44e-11 & 5.72e-04 & 5.72e-09 & 5.73e-09 \\
    G5 & HRS5 & -1.1 & 0.1 & 1.88e-06 & 1.00e-05 & 1.88e-11 & 6.30e-04 & 6.30e-09 & 6.32e-09 \\
    G6 & HRS6 & -1.0 & 0.1 & 2.57e-06 & 1.00e-05 & 2.57e-11 & 7.06e-04 & 7.06e-09 & 7.09e-09 \\
    G7 & HRS7 & -0.9 & 0.1 & 3.80e-06 & 1.00e-05 & 3.80e-11 & 7.76e-04 & 7.76e-09 & 7.80e-09 \\
    G8 & LRS  & -0.5 & 0.1 & 5.09e-06 & 1.00e-05 & 5.09e-11 & 3.97e-04 & 3.97e-09 & 4.02e-09 \\
    \bottomrule
    \end{tabular}
\end{sidewaystable}

\begin{sidewaystable}
    \small
    \setlength{\tabcolsep}{5.0pt}
    \renewcommand{\arraystretch}{2.0}
    \centering
    \caption{\textbar\quad Energy consumption (\textit{PtHT}). Measured parameters for various conductance states (G1--G8).}
    \label{tab:ptht_energy}
    \begin{tabular}{
        p{0.8cm}   
        p{1.5cm}   
        C{1.5cm}     
        C{1.5cm}   
        C{1.5cm}   
        C{1.5cm}     
        C{1.5cm}     
        C{1.5cm}     
        C{2.0cm}     
        C{1.5cm}     
        C{1.3cm}     
    }
    \toprule
      & & & & & & & & \multicolumn{2}{c}{\textbf{Averaged Power and Energy}} & \\
    \cmidrule(lr){9-10}
      \textbf{G} & \textbf{R\_states} & 
      \(\mathbf{V_{\text{RESET}}}\)(V) & 
      \(\mathbf{V_{\text{SET}}}\)(V) &
      \(\mathbf{V_{\text{read}}}\)(V) &
      \(\mathbf{P_{\text{read}}}\)(W) & 
      \(\mathbf{t_{\text{read}}}\)(s) & 
      \(\mathbf{E_{\text{read}}}\)(J) &
      \(\mathbf{P_{\text{program}}}\)(W) & 
      \(\mathbf{E_{\text{program}}}\)(J) & 
      \(\mathbf{E_{\text{tot.}}}\)(J) \\
    \midrule
      G1 & HRS1 & -1.5 & 1.1 & 0.1 & 5.96e-06 & 1.00e-05 & 5.96e-11 & 8.92e-04 & 8.92e-09 & 8.98e-09 \\
      G2 & HRS2 & -1.4 & 1.1 & 0.1 & 6.27e-06 & 1.00e-05 & 6.27e-11 & 9.44e-04 & 9.44e-09 & 9.51e-09 \\
      G3 & HRS3 & -1.3 & 1.1 & 0.1 & 6.69e-06 & 1.00e-05 & 6.69e-11 & 1.00e-03 & 1.00e-08 & 1.01e-08 \\
      G4 & HRS4 & -1.2 & 1.1 & 0.1 & 7.35e-06 & 1.00e-05 & 7.35e-11 & 1.07e-03 & 1.07e-08 & 1.08e-08 \\
      G5 & HRS5 & -1.1 & 1.1 & 0.1 & 8.39e-06 & 1.00e-05 & 8.39e-11 & 1.12e-03 & 1.12e-08 & 1.13e-08 \\
      G6 & HRS6 & -1.0 & 1.1 & 0.1 & 8.95e-06 & 1.00e-05 & 8.95e-11 & 1.07e-03 & 1.07e-08 & 1.07e-08 \\
      G7 & HRS7 & -0.9 & 1.1 & 0.1 & 9.02e-06 & 1.00e-05 & 9.02e-11 & 9.78e-04 & 9.78e-09 & 9.87e-09 \\
      G8 & LRS  & -0.5 & 1.1 & 0.1 & 4.51e-06 & 1.00e-05 & 4.51e-11 & 5.88e-04 & 5.88e-09 & 5.93e-09 \\
    \bottomrule
    \end{tabular}
\end{sidewaystable}

\begin{table}
    \small
    \setlength{\tabcolsep}{5pt}
    \renewcommand{\arraystretch}{2.0}
    \centering

    \caption{\textbar\quad Write energy of one-time programming of synaptic weights for the pre-trained \textit{N-MNIST SNN}.}
    \label{tab:nmnist-energy}

    \begin{tabular}{l c c c c c}
    \toprule
      & & \multicolumn{2}{c}{\textbf{Write Energy (J)}} & 
          \multicolumn{2}{c}{\textbf{Average Write Energy per Synapse (J)}} \\
    \cmidrule(lr){3-4}\cmidrule(lr){5-6}
      \textbf{Layer} & \textbf{Synapses \#} &
      \textbf{PtHT} & \textbf{PdHT} &
      \textbf{PtHT} & \textbf{PdHT} \\
    \midrule
    SC1 & 588   & 1.57e-05 & 9.15e-06 & 2.67e-08 & 1.56e-08 \\
    SC2 & 2400  & 6.31e-05 & 3.68e-05 & 2.63e-08 & 1.54e-08 \\
    SC3 & 48000 & 1.34e-03 & 7.69e-04 & 2.78e-08 & 1.60e-08 \\
    SF1 & 10080 & 2.79e-04 & 1.62e-04 & 2.77e-08 & 1.61e-08 \\
    SF2 & 840   & 2.22e-05 & 1.40e-05 & 2.64e-08 & 1.67e-08 \\
    \midrule
    Total & 61908 & 1.72e-03 & 9.91e-04 & 2.77e-08 & 1.60e-08 \\
    \bottomrule
    \end{tabular}
\end{table}

\begin{table}
    \small
    \setlength{\tabcolsep}{5pt}
    \renewcommand{\arraystretch}{2.0}
    \centering
    \caption{\textbar\quad Write energy of one-time programming of synaptic weights for the pre-trained Gesture \textit{SNN}.}
    \label{tab:gesture-energy}

    \begin{tabular}{l c c c c c}
    \toprule
      & & \multicolumn{2}{c}{\textbf{Write Energy (J)}} & 
          \multicolumn{2}{c}{\textbf{Average Write Energy per Synapse (J)}} \\
    \cmidrule(lr){3-4}\cmidrule(lr){5-6}
      \textbf{Layer} & \textbf{Synapses \#} & 
      \textbf{PtHT} & \textbf{PdHT} &
      \textbf{PtHT} & \textbf{PdHT} \\
    \midrule
    SC1   & 800      & 2.15e-05 & 1.25e-05 & 2.69e-08 & 1.57e-08 \\
    SC2   & 4608     & 1.21e-04 & 7.06e-05 & 2.63e-08 & 1.53e-08 \\
    SF1   & 1048576  & 2.91e-02 & 1.66e-02 & 2.77e-08 & 1.58e-08 \\
    SF2   & 5632     & 1.55e-04 & 8.86e-05 & 2.76e-08 & 1.57e-08 \\
    \midrule
    Total & 1059616  & 2.94e-02 & 1.68e-02 & 2.77e-08 & 1.58e-08 \\
    \bottomrule
    \end{tabular}
\end{table}

\begin{table}
    \small
    \setlength{\tabcolsep}{5pt}
    \renewcommand{\arraystretch}{2.0}
    \centering
    \caption{\textbar\quad Read energy per layer during inference of the \textit{N-MNIST SNN}.}
    \label{tab:nmnist-read-energy}
    \begin{tabular}{l c c c c c}
      \toprule
       & & \multicolumn{2}{c}{\textbf{Read Energy (J)}} &
           \multicolumn{2}{c}{\textbf{Average Read Energy per Synapse (J)}} \\
      \cmidrule(lr){3-4}\cmidrule(lr){5-6}
      \textbf{Layer} & \textbf{Synapses \#} &
      \textbf{PtHT} & \textbf{PdHT} &
      \textbf{PtHT} & \textbf{PdHT} \\
      \midrule
      SC1  & 588    & 2.61e-04 & 7.36e-05 & 4.43e-07 & 1.25e-07 \\
      SC2  & 2400   & 8.04e-03 & 2.29e-03 & 3.35e-06 & 9.55e-07 \\
      SC3  & 48000  & 2.17e-03 & 5.55e-04 & 4.52e-08 & 1.16e-08 \\
      SF1  & 10080  & 3.65e-04 & 9.93e-05 & 3.62e-08 & 9.83e-09 \\
      SF2  & 840    & 1.82e-05 & 5.85e-06 & 2.17e-08 & 6.96e-09 \\
      \midrule
      Total & 61908 & 1.08e-02 & 3.03e-03 & 1.75e-07 & 4.89e-08 \\
      \bottomrule
    \end{tabular}
\end{table}

\begin{table}
    \small
    \setlength{\tabcolsep}{5pt}
    \renewcommand{\arraystretch}{2.0}
    \centering
    \caption{\textbar\quad Read energy per layer during inference of the Gesture \textit{SNN}.}
    \label{tab:gesture-read-energy}
    \begin{tabular}{l c c c c c}
      \toprule
       & & \multicolumn{2}{c}{\textbf{Read Energy (J)}} &
           \multicolumn{2}{c}{\textbf{Average Read Energy per Synapse (J)}} \\
      \cmidrule(lr){3-4}\cmidrule(lr){5-6}
      \textbf{Layer } & \textbf{Synapses \#} &
      \textbf{PtHT} & \textbf{PdHT} &
      \textbf{PtHT} & \textbf{PdHT} \\
      \midrule
      SC1  & 800      & 9.93e-03 & 2.63e-03 & 1.24e-05 & 3.29e-06 \\
      SC2  & 4608     & 3.23e-02 & 9.23e-03 & 7.01e-06 & 2.00e-06 \\
      SF1  & 1048576  & 4.69e-02 & 1.23e-02 & 4.48e-08 & 1.17e-08 \\
      SF2  & 5632     & 4.85e-05 & 1.58e-05 & 8.61e-09 & 2.81e-09 \\
      \midrule
      Total & 1059616 & 8.92e-02 & 2.42e-02 & 8.42e-08 & 2.28e-08 \\
      \bottomrule
    \end{tabular}
\end{table}

\begin{figure}[!ht]
\centerline{\includegraphics[width=0.6\linewidth]{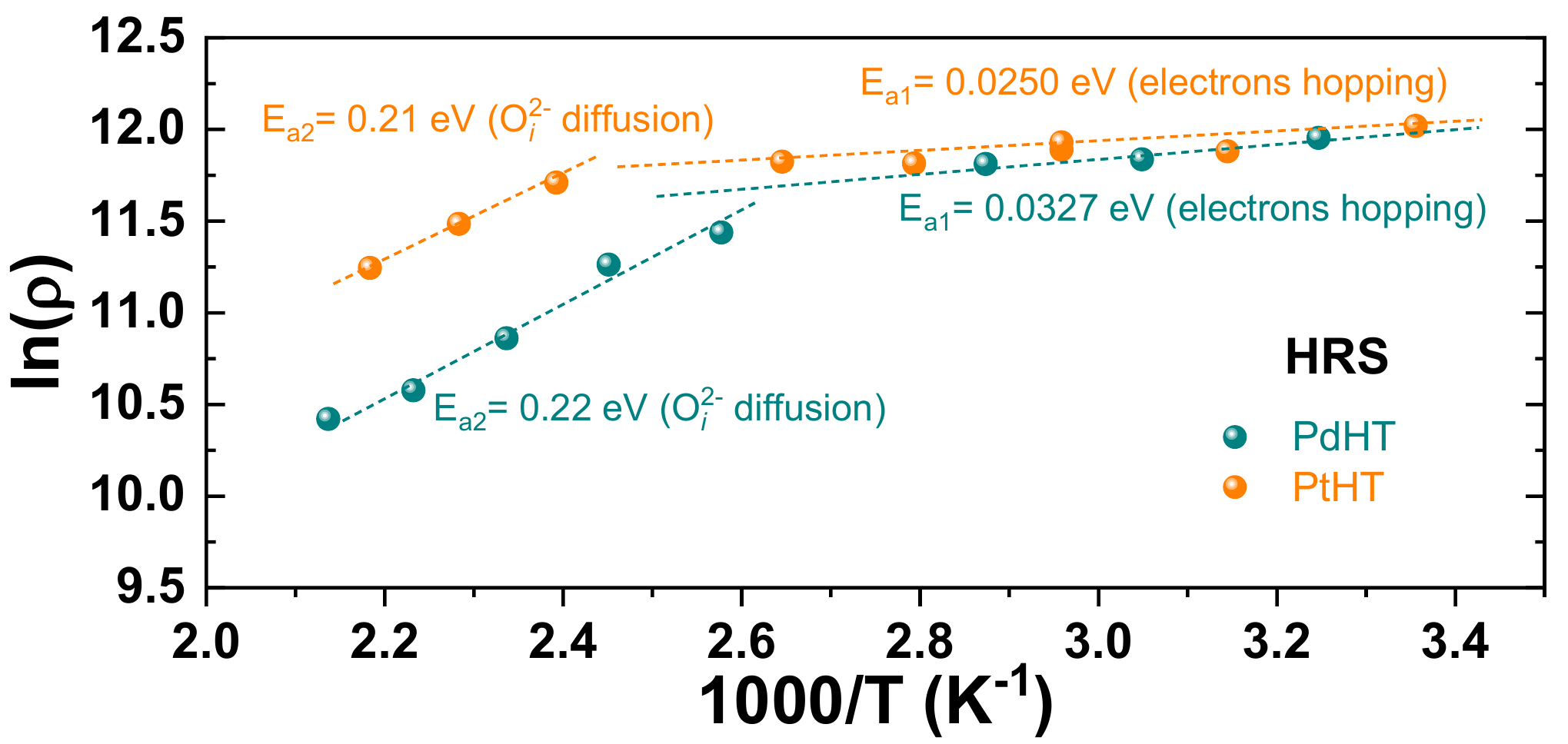}}
\caption{\textbar\quad \textbf{Activation energy comparison for PdHT and PtHT}} 
\label{esi_ea}
\vspace{-1em}%
\end{figure}

\begin{figure}[!ht]
\centerline{\includegraphics[width=0.6\linewidth]{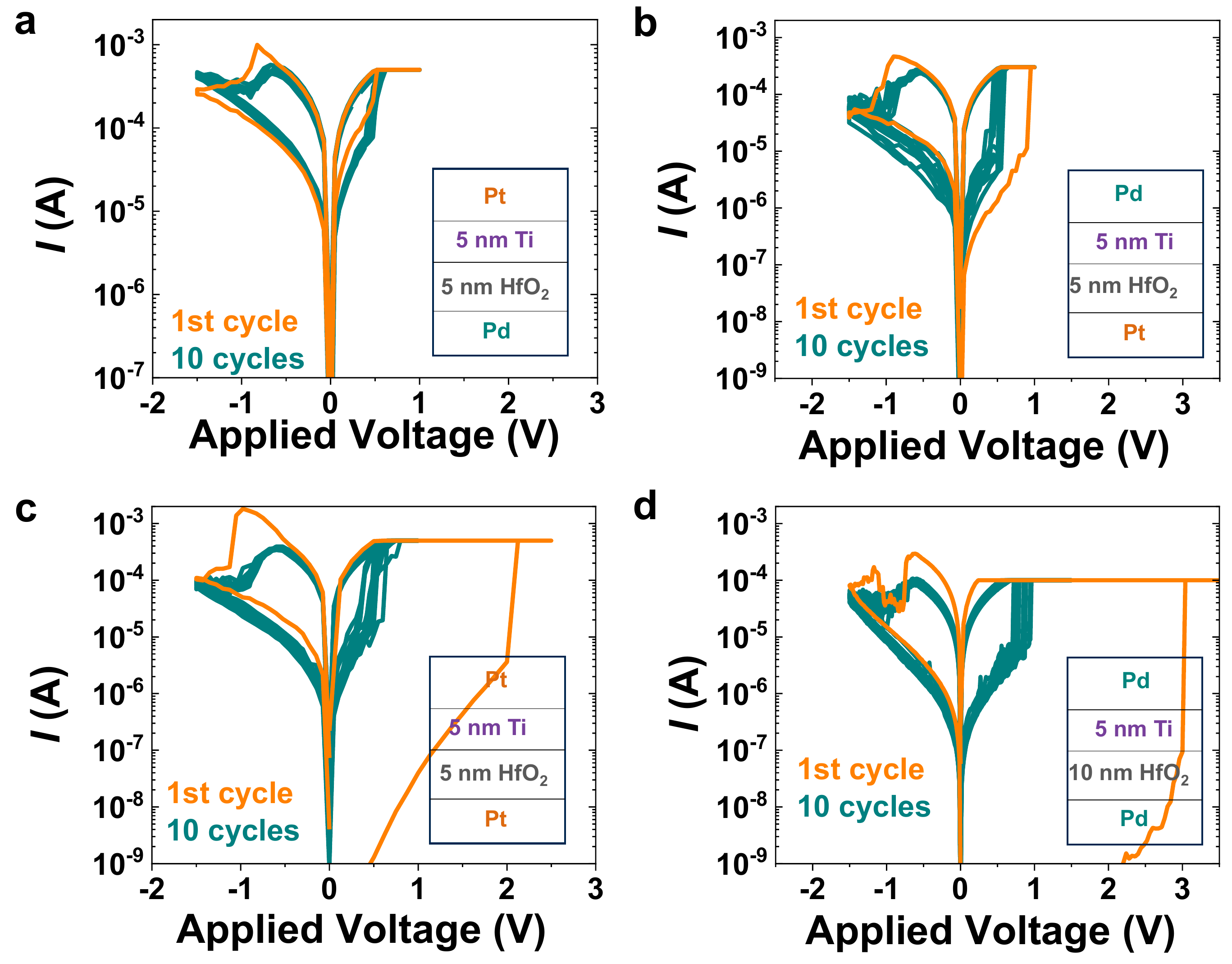}}
\caption{\textbar\quad\ Different combination of Pt and Pd as \textit{BE/TE} and their I-V curve with 5nm of HfO\textsubscript{2-x} (a, b, c) and 5nm of HfO\textsubscript{2-x} with Pd electrodes for \textit{BE/TE} (d). } 
\label{esi_electrodes}
\vspace{-1em}%
\end{figure}

\begin{table}[!ht]
\caption{\textbar\quad Average read energy per spike.}
\centering
\begin{tabular}{lcc}
\toprule
      & N-MNIST SNN & Gesture SNN \\
\midrule
PtHT   & 33.6\,nJ    & 65.7\,nJ     \\
PdHT   & 9.4\,nJ     & 17.7\,nJ     \\
\bottomrule
\end{tabular}
\label{tab:read_energy_spike}
\end{table}

\end{document}